\documentclass{aa}
\usepackage{graphicx}
\usepackage{txfonts}
\begin{document}

   \title{The bi-modal $^7$Li distribution of the Milky Way's thin-disk dwarf stars}

   \subtitle{The role of Galactic-scale events and stellar evolution}
\titlerunning{The A(Li) bi-modality and the Galaxy evolution}
\authorrunning{Roca-F\`abrega et al.}
   \author{S. Roca-F\`abrega
          \inst{1}\fnmsep\thanks{sroca01@ucm.es},
          F. Llorente de Andr\'es
          \inst{2,3},
          C. Chavero
          \inst{4},
          C. Cifuentes
          \inst{3}
          \and
          R. de la Reza\inst{5}
          }

   \institute{Departamento de F\'isica de la Tierra y Astrof\'isica and IPARCOS, Facultad de Ciencias F\'isicas, Plaza Ciencias, 1, Madrid, E-28040, Spain
   \and
   Ateneo de Almagro, Secci\'on de Ciencia y Tecnolog\'ia, 13270 Almagro, Spain
   \and
   Departmento de Astrof\'isica, Centro de Astrobiolog\'ia (CAB, CSIC-INTA), ESAC Campus, Camino Bajo del Castillo s/n, 28692 Villanueva de la Cañada, Madrid, Spain
   \and
   Observatorio Astron\'omico de C\'ordoba, Universidad Nacional de C\'ordoba, Laprida 854, 5000 C\'ordoba, CONICET, Argentina
   \and
   Observatório Nacional, Rua General Jos\'e Cristino 77, 28921-400 São Cristovão, Rio de Janeiro, RJ, Brasil}

   \date{Received XXXX; accepted XXXX}

 
  \abstract
   {The lithium abundance, A(Li),  in stellar atmospheres suffers from various enhancement and depletion processes during the star’s lifetime. While several studies have demonstrated that these processes are linked to the physics of stellar formation and evolution, the role that Galactic-scale events play in the galactic A(Li) evolution is not yet well understood.}
   {We aim to demonstrate that the observed A(Li) bi-modal distribution, in particular in the FGK-dwarf population, is not a statistical artefact and that the two populations connect through a region with a low number of stars. We also want to investigate the role that Galactic-scale events play in shaping the A(Li) distribution of stars in the thin disk.}
   {We use statistical techniques along with a Galactic chemical evolution model for A(Li) that includes most of the well-known $^7$Li production and depletion channels.}
   {We confirm that the FGK main-sequence stars belonging to the Milky Way’s thin disk present a bi-modal A(Li) distribution. We demonstrate that this bi-modality  can be generated by a particular Milky Way star formation history profile combined with the stellar evolution’s $^7$Li depletion mechanisms. We show that A(Li) evolution can be used as an additional proxy for the star formation history of our Galaxy.}
   {}


   \maketitle
%

\section{Introduction}
The evolution of the $^7$Li abundance, A(Li), in stellar atmospheres as a function of their age has challenged researchers since the proposal of the `cosmological lithium problem' \citep[e.g.][]{Fields2011}. The discrepancy between the expected primordial A(Li)  \citep{PlanckCollaboration2016} and the one measured in population II stars \citep[Spite plateau in][]{Spite2012, Tanabashi2018}, which is a factor of three to four lower, is still difficult to explain. Furthermore, the A(Li) of young T-Tauri stars was shown to be almost a factor of ten higher than that of Population II stars \citep{Frasca2018} and 2$-$3\,dex higher than that of Sun-like stars \citep{LopezValdivia2015}. This last result added a new level of complexity to the lithium problem, now referred as the `Galactic lithium problem'. \smallskip
The complexity of this problem resides in the variety of mechanisms that rule the evolution of individual stars. In particular, two different physical processes compete on the creation and destruction of lithium in stars: (i) $^7$Li and $^6$Li ions are destroyed and diluted in the epidermal stellar layers of most FGK-dwarf stars, although they can be partially preserved in the atmosphere of some metal-rich, warm F dwarfs \citep[e.g.][]{Charbonnel2021}, and (ii) new $^7$Li can be created in stellar interiors and transported to the stellar atmosphere.

The Li depletion occurs all along the stellar evolution in regions where the temperature exceeds $\sim$2.5~10$^6$~K, from the pre-main sequence (PMS), to the main sequence (MS), and up to the first dredge-up in the initial red giant branch (RGB) stages. It is in the PMS stage that the most important initial Li depletion occurs \citep[see for example][]{Eggenberger2012}. Several authors have presented a deep analysis of Li depletion mechanisms in stars (see for example \citealt{Chavero2019}, \citealt{ArencibiaSilva2020}, \citealt{Cassisi2020}, \citealt{Dumont2021}, and references therein).
The Li creation occurs in the RGB stage and in the first red clump (RC) stage. Here, a complex process involves the transport of $^7$Be from the stellar interior to the stellar surface. In this process, $^7$Be decays into fresh $^7$Li at the surface. In fact, very large abundances can be found in the upper part of the RGB stars and, especially, in the initial RC, where several authors have reported giants that are more than three times Li-richer than in the RGB \citep[see for instance][]{Martell2020}. A(Li) values as high as 4.5 have been observed and studied by \citet{Yan2018} and \citet{delaReza2020} in evolved stars.\smallskip

Additionally, a $^7$Li enhancement has also been observed in young stars; this result requires new theories to be understood. Theories with a larger degree of success in explaining the A(Li) enhancement in the Galaxy have focused on the Galactic chemical evolution \citep[GCE; e.g.][]{Grisoni2019,CescuttiMolaro2019}. These models predict an increase in the A(Li) that follows the pollution of interstellar gas metals by  RGB stars \citep{CameronFowler1971}, novae \citep{Romano2001,Starrfield2020}, supernovae \citep[SNe;][]{Matteucci2006}, and  asymptotic giant branch (AGB) stars \citep{Matteucci2014}. The aforementioned increase in metallicity has a strong impact on metallicity-dependent $^7$Li production and depletion mechanisms, such as the properties and size of protostellar disks \citep[e.g.][]{Machida2015}. Planet formation and evolution may also play an important role in the A(Li) evolution of the host star. In particular, the effect that tidal interactions with the protoplanetary disks or young planets  -- and their consequent engulfment -- have on the stellar A(Li) is still under study. Our team is already tackling this problem, and our results will be presented in a forthcoming paper. Finally, some authors have also proposed that physical mechanisms that have a low or no relation with stellar evolution can also have an impact on the interstellar A(Li) evolution. Some examples are the destruction or production of lithium ions by energetic cosmic rays (CRs) in Galactic gas clouds  \citep{Fields1994} or changes in the interstellar gas metallicity due to low-metallicity gas inflows. \smallskip

In the light of these results and of the growing complexity of the `A(Li) problem', the need for a global theory to explain A(Li) evolution is evident. This theory needs to include not only a deep understanding of stellar evolution and planet formation, but also of galactic formation and evolution. \smallskip

In the past, our knowledge of Milky Way (MW) formation and evolution was limited to a few theoretical models based on observations of external galaxies and a few observations of stars in the solar neighbourhood. In the last decade, a flood of high quality data of physical properties and positions of stars in our Galaxy has allowed researchers to better constrain its recent history \citep{Haywood2013,Cutri2013,DR2Gaia2018,Helmi2018,Mor2019,EDR3Gaia2021}. Currently, we have a much better understanding of the recent MW star formation history (SFH) and its chemodynamical evolution \citep{Haywood2016,RuizLara2020}. Using all this new knowledge, we are now ready to investigate the A(Li) evolution from a global perspective.\smallskip

This is a new paper within our collaboration on the study of the A(Li) evolution of Population I stars in the solar neighbourhood (see \citealt{Chavero2019} and \citealt{Llorente2021}, hereafter CH19 and LA21, respectively). Here, we go one step further in our analysis, studying the effect that Galactic evolution has on the thin-disk stars' A(Li) distribution. In forthcoming papers of this collaboration we will present results on the study of planet formation and evolution and their relation with Galactic-scale events and A(Li) evolution, as well as results on stellar kinematics, both of the thin and the thick disk, and its relation with the galactic open clusters (OCs) and moving groups (MGs).\\

This paper is organised as follows.
In Sect.~\ref{sec:observations} we describe the observational samples. We present our A(Li) chemical evolution model in Sect.~\ref{sec:theory}. A detailed analysis of the presence of two A(Li) populations (rich and poor) in the thin-disk stars, and their connection to each other through what we called `the isthmus', is presented in Sect.~\ref{sec:isthmus}. Section~\ref{sec:galactic} is devoted to establishing a link between the theory of galaxy formation and evolution, observations of the MW’s SFH, and the A(Li) evolution of thin-disk stars. In Sect.~\ref{sec:conc} we summarise the main conclusions of this work and give a preview of the future work of this collaboration. \smallskip

\section{Observational sample}\label{sec:observations}

In this section we describe the data sample used in this work. The full procedure is described in LA21.\\

\subsection{Stellar parameters and catalogue description}

The main sample used in this work is an update of the one used in \citet{Chavero2019}, with a significant increase in the number of stars with good A(Li) measurements. The A(Li) values were obtained from 20 catalogues (see Table 1 in LA21). All catalogues were cross-matched, and the properties of each star were carefully studied to obtain a homogeneous sample of field FGK MS stars. When more than one value for a specific property of a star was available (i.e. from different catalogues), we took, preferentially, the one from CH19. When not available in CH19, we chose the corresponding value from the most recent publication. By following this process, the $T_{\rm eff}$ and parallax were mainly obtained from the sub-sample of {\em Gaia} Data Release 2 (DR2) stars \citep{DR2Gaia2018} recommended by \citet{Andrae2018}. Similarly, most of the [Fe/H] values were collected from the work by \citet{Gaspar2016}.

A relevant variable in the current work is age. In order to get a better picture of the A(Li) evolution with age, and to reduce systematics, we decided to use two independent age scales. First, we obtained age values from the literature, in particular from \citet{DelgadoMena2015} (see LA21 for a full list of used catalogues), and we named those {\it Age\_Lite}. The second age scale was obtained by using the {\tt Param1.3}\footnote{http://stev.oapd.inaf.it/cgi-bin/param\_1.3} code \citep{daSilva2006}, and we named it {\it Age\_Param}. This code requires T$_{\rm eff}$, [Fe/H], $V$ magnitude, and parallax values.
Our final sample contains 1385 stars with reliable A(Li) values.

Some studies have suggested that the presence of planets can have an impact on the A(Li) evolution of stars \citep{AguileraGomez2016,Stephan2020,SoaresFurtado2020}. After carefully revising the NASA Exoplanet Archive\footnote{\url{https://exoplanetarchive.ipac.caltech.edu/}}, we obtained a sub-sample of 289 stars with confirmed exoplanets. In the following sections we refer to the sub-sample of 289 stars with confirmed exoplanets as Y stars and the remaining as N stars (i.e. stars without detected planets).\\

Additionally, we used data from the third data release (DR3) of the Gaia ESO Survey \citep[GES;][]{Gilmore2012}\footnote{http://www.eso.org/rm/api/v1/public/releaseDescriptions/91} and from the Galah-DR3 catalogue \citep{Buder2021}. We selected only FGK-dwarf stars (4000$<$T$_{\rm eff}<$7000~K and 3.6$<$logg$<$4.6) with well-derived A(Li), obtaining two sub-samples of 2337 and 275612 stars, respectively. After a deep analysis, we found that the subset of stars with well-derived A(Li) is biased against cool PMS and MS FGK-dwarf stars \citep[e.g. Sect. 4.1.1 in][]{Gao2018} and in GES also biased in favour of stars belonging to young stellar associations. We concluded that these catalogues are not useful for our study of the field FGK-dwarf star A(Li) distribution. Consequently, in this work we focus our analysis on the sample of stars presented by LA21.

\subsection{Sample characterisation}

In order to avoid any bias due to the data collection, and to show that the results are not statistically dependent on our sample, we compared values within it with those in three published samples of field stars: \citet{AguileraGomez2018}, \citet[][the AMBRE project]{Guiglion2016}, and \citet{BensbyLind2018}, hereafter AG18, Gu16, and BL18, respectively. It is relevant to mention that the field stars in AG18 were collected from different sources (see references therein) and well characterised afterwards. Our catalogue is similar to the one from AG18 but with a lower number of 11$-$12\,Gyr stars. This is because in our work we are interested in thin-disk stars, while AG18 include halo K-type dwarf stars. We also reviewed the large catalogue from Gu16, the AMBRE project. An important difference between our sample and the Gu16 samples is that in Gu16 the authors took the maximum of the projected rotation speed (between 10 and 15\,kms$^{-1}$) to derive A(Li) values, while we did not impose any restriction. The young stars usually show higher rotation speeds, so we expect to see a bias towards young stars when comparing the A(Li) values presented by Gu16 with ours. Finally, we also considered the catalogue published by BL18, which is similar to ours but includes stars with ages of 10\,Gyr and above and with globally lower metallicities.
After carefully studying the A(Li) values and the A(Li)-[Fe/H] relation in all four catalogues, we conclude that our sample shows similar distributions, with no evident bias apart from the one expected in the Gu16 sample. We also observed only minor differences in the dispersion of the A(Li)-[Fe/H] relation. 

\subsection{Kinematics and galactic component decomposition}\label{sec:data_kin}

\begin{figure}
    \centering
    \includegraphics[width=0.45\textwidth]{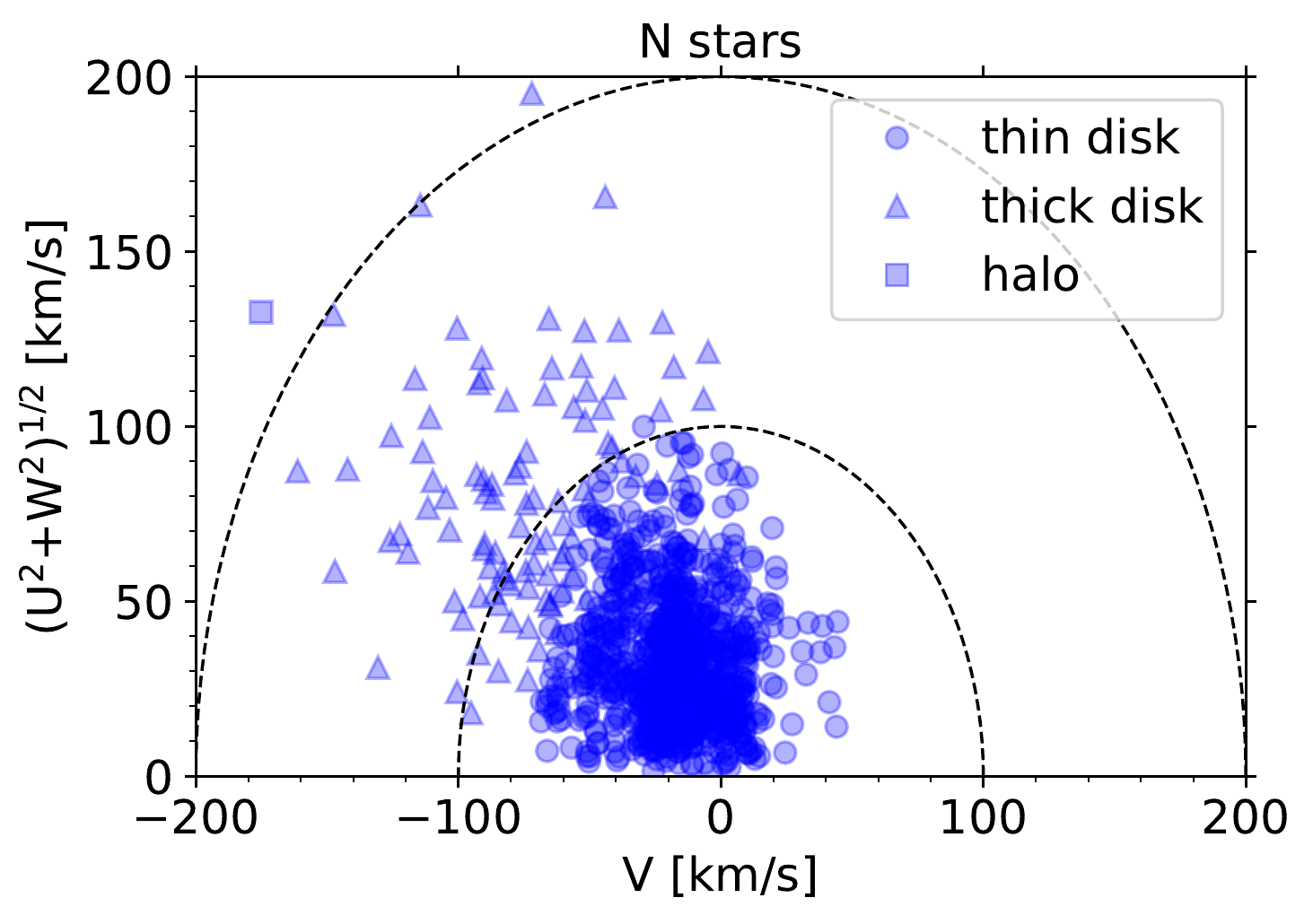}
     \includegraphics[width=0.45\textwidth]{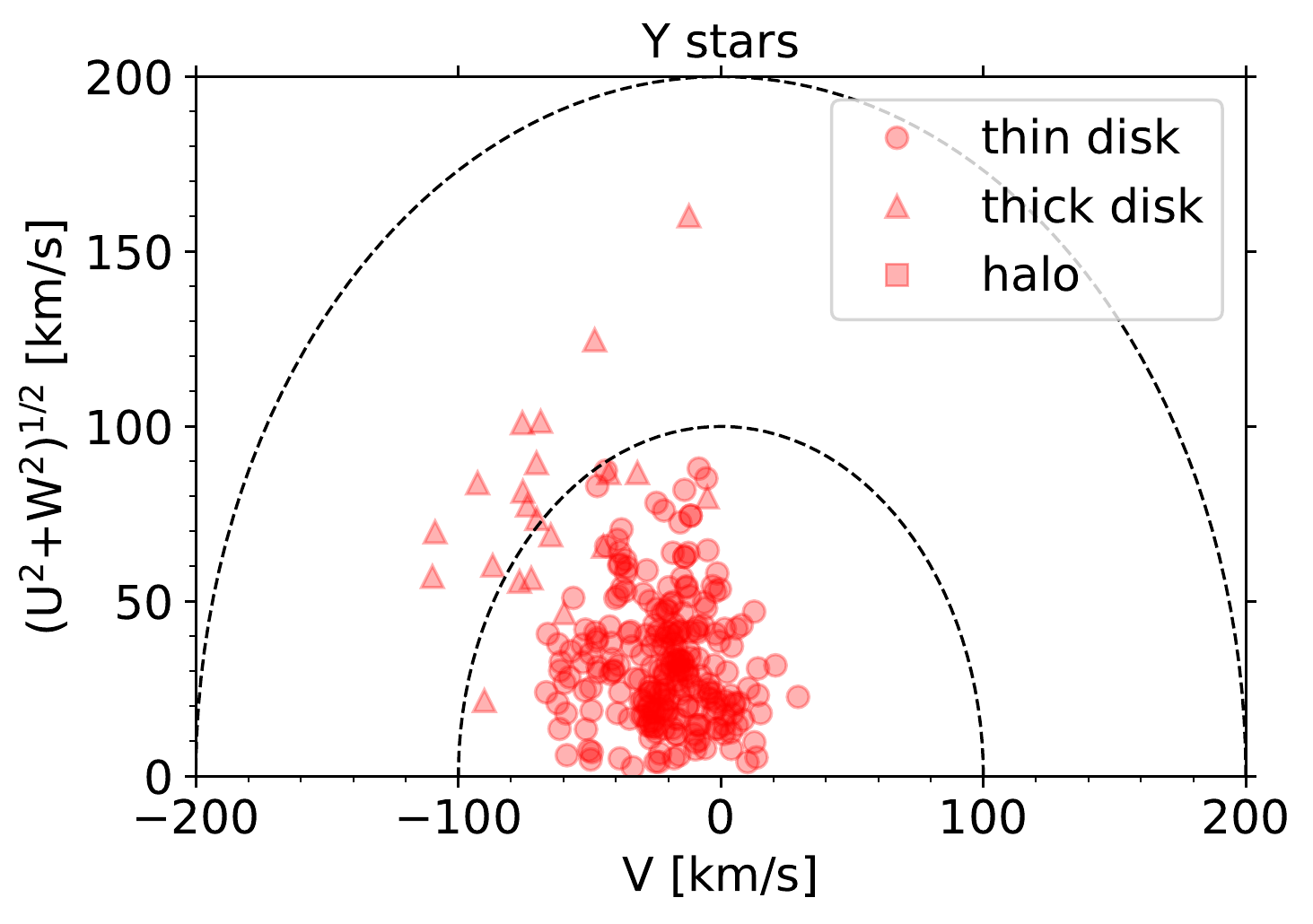}
    \caption{Toomre diagrams \citep[see Fig. 9 in][]{Kovalev2019} of stars without planets (top panel) and with planets (bottom panel) in the LA21 sample. We show thin-disk stars as circles, thick-disk stars as triangles, and halo stars as squares. We followed the kinematic decomposition proposed by \citet{Bensby2014}.}
    \label{fig:Toomre}
\end{figure}

In addition to the stellar parameters mentioned above, we used parallaxes, proper motions, and radial velocities from {\em Gaia} DR2 \citep{DR2Gaia2018} to obtain the galactocentric velocities $U,V$, and $W$, using the public software package {\tt PYGAIA}\footnote{https://github.com/agabrown/PyGaia}. Using this kinematic information, we constructed Toomre diagrams (Fig.~\ref{fig:Toomre}) and kinematically decomposed our samples into three galactic components: stars from the thin disk, stars from the thick disk, and stars from the galactic halo. For the decomposition we followed the strategy presented in \citet{Bensby2014}. Figure~\ref{fig:Toomre} shows that most of the stars in our main sample belong to the thin-disk galactic component.  This result is not surprising, as most of the stars in this work are located in the solar neighbourhood.

This kinematic decomposition is relevant for the current work because each galactic population (i.e. thin disk, thick disk, and halo) undergoes a different chemical evolution. Using a single galactic population reduces the number of variables to analyse when interpreting the A(Li) and [Fe/H] evolution. 
After selecting only thin-disk stars, our final sample contains 1298 N stars and 267 Y stars.

\section{Galactic A(Li) evolution model}\label{sec:theory}
In this section we present our GCE model. We developed this model to reproduce the A(Li) evolution of thin-disk stars. Here we summarise the theory of A(Li) evolution that supports the assumptions and approximations that we applied in our algorithms, we describe the included physical processes, and, finally, we present our default models for the thin-disk stars' A(Li) evolution.

\subsection{The A(Li) evolution in stars and in the interstellar medium gas} \label{subsec:lithiumISM}

The A(Li) evolution in the Universe is a complex process as it is strongly connected with both Galactic-scale events and stellar evolution \citep[see][]{Lyubimkov2016,Randich2021}.

According to several observations and analytical models \citep[e.g.][and references therein]{Steigman2010}, the initial A(Li) from the primordial nucleosynthesis is A(Li)=2.7\,dex. If no other mechanism of $^7$Li production or depletion were present in the intergalactic medium (IGM), this value would correspond to the primordial IGM gas and as such to the initial A(Li) of newborn stars. This is not the case in most Population I and II stars.

From the proto-stellar phase, several depletion mechanisms begin to play a role in modifying the A(Li) abundance in stellar atmospheres. There is an extensive literature showing that old stars in the halo, in globular clusters, and in some dwarf galaxies \citep[e.g. Sagittarius dwarf, Gaia-Enceladus in][]{Molaro2020} show a A(Li) plateau at $\sim{\rm A(Li)}=2.2$, which is 0.5\,dex lower than the primordial value \citep[e.g.][]{Spite2012} known as the `Spite plateau'. It is still unclear how this decrease in the A(Li) occurs and what its dependence on the initial stellar metallicity is. In fact, some old dwarf stars in the halo show A(Li) far below the plateau \citep{Randich2010}. Some authors have suggested that a difference may exist between the early depletion mechanism (PMS) in metal-rich stars and in metal-poor stars, probably related with the different mass, dynamics, and age of the young stellar accretion disk. One example is the Eggenberger depletion mechanism \citep{Eggenberger2012} that sets a correlation between $^7$Li depletion and the stellar rotation. The initial rotation speed is a parameter that depends on the size and lifetime of the stellar accretion disk, variables that have been shown to be strongly correlated with stellar metallicity (CH19). \citet{Eggenberger2012} and CH19 demonstrated that fast rotators can only deplete the lithium down to the Spite plateau (A(Li)$\sim 2.2$), while slow rotators can do so down to zero \citep[see][]{ArencibiaSilva2020}. Although further investigations are required, this result gives some hints as to the origin of the observed differences between the A(Li) in halo metal-poor stars and in the thin-disk metal-rich stars.

Surprisingly, when observing young stars in the thin disk, it was found that the initial A(Li) is about $\log{\rm A(Li)}=3.2$, which is ten times higher than the aforementioned Spite plateau. Which mechanisms dominate in this $^7$Li enhancement is still a matter of discussion, but the most recent works point towards pollution from ongoing stellar evolution (i.e. stellar feedback) and, mostly, from low-mass stars (see \citealt{Romano2001}, \citealt{Matteucci2014}, and a detailed discussion in \citealt{CescuttiMolaro2019}). Using galactic chemical models, several authors have concluded that about two-thirds of the current A(Li) over-abundance in the interstellar medium (ISM) was produced by (i) winds from red giants with masses from 1 $\mathcal{M}_\odot$ to 2 $\mathcal{M}_\odot$ -- low-intermediate-mass stars (LIMSs) -- that synthesised $^7$Li via the Cameron-Fowler mechanism \citep{CameronFowler1971}, and (ii) by novae \citep{CescuttiMolaro2019}. More recently, several authors have proposed that CRs can play an important role in $^7$Li levels through the spallation of more massive atoms \citep{Smiljanic2009,Grisoni2019}.
On the other hand, most works agree that core-collapse SNe (high-mass stars) and type-Ia SNe (SNeIa) probably contributed to less than 10\%. In this context, it is expected that the ISM pollution with A(Li)-enhanced gas is a slow but continuous process that starts after strong star formation events, mainly due to low-mass star feedback, spanning up to $\sim 5$\,Gyr from the star formation event \citep{Cescutti2008,CescuttiMolaro2019}. High-mass stars explode as core-collapse SNe, releasing a vast amount of energy and metals into the ISM, shortly after star formation \citep{Romano2005,Cescutti2008}; although atmospheres of high-mass stars are highly A(Li) depleted \citep{Lambert1980}, the production of $^7$Li via the $\nu$ process \citep{Nakamura2010} prevents the reduction of A(Li) in the ISM gas by these type-II SN (SNII) events.

In the previous paragraphs we summarised the mechanisms that trigger changes in the A(Li) evolution in stars and ISM gas. However, A(Li) data mostly come from stellar atmospheres, not from direct observations of the ISM gas abundances. Therefore, to tackle the Galactic A(Li) from its value in stellar atmospheres, we need to account not only for the ongoing depletion and production  mechanisms during all the stars' lifetimes \citep{CescuttiMolaro2019}, but also the Galactic-scale events that change the A(Li) of the ISM gas A(Li). This because it is the A(Li) in the ISM gas that the newborn stars inherit.\\

\subsection{The model}\label{sec:Chem_model}

Our GCE model includes most of the well-known mechanisms of $^7$Li production and destruction inside and outside stars, in agreement with previous works such as \citet{Grisoni2019} and \citet{Dumont2021}. It is important to note that with this model we do not seek a prediction of the A(Li) evolution in individual stars but rather the bulk evolution in the field FGK-dwarf thin-disk population. We built the model following the prescriptions presented by \citet{CescuttiMolaro2019} and \citet{Grisoni2019}, but adding new physical processes and making small corrections to the former recipes (described below).

\begin{figure}
    \centering
    \includegraphics[width=0.45\textwidth]{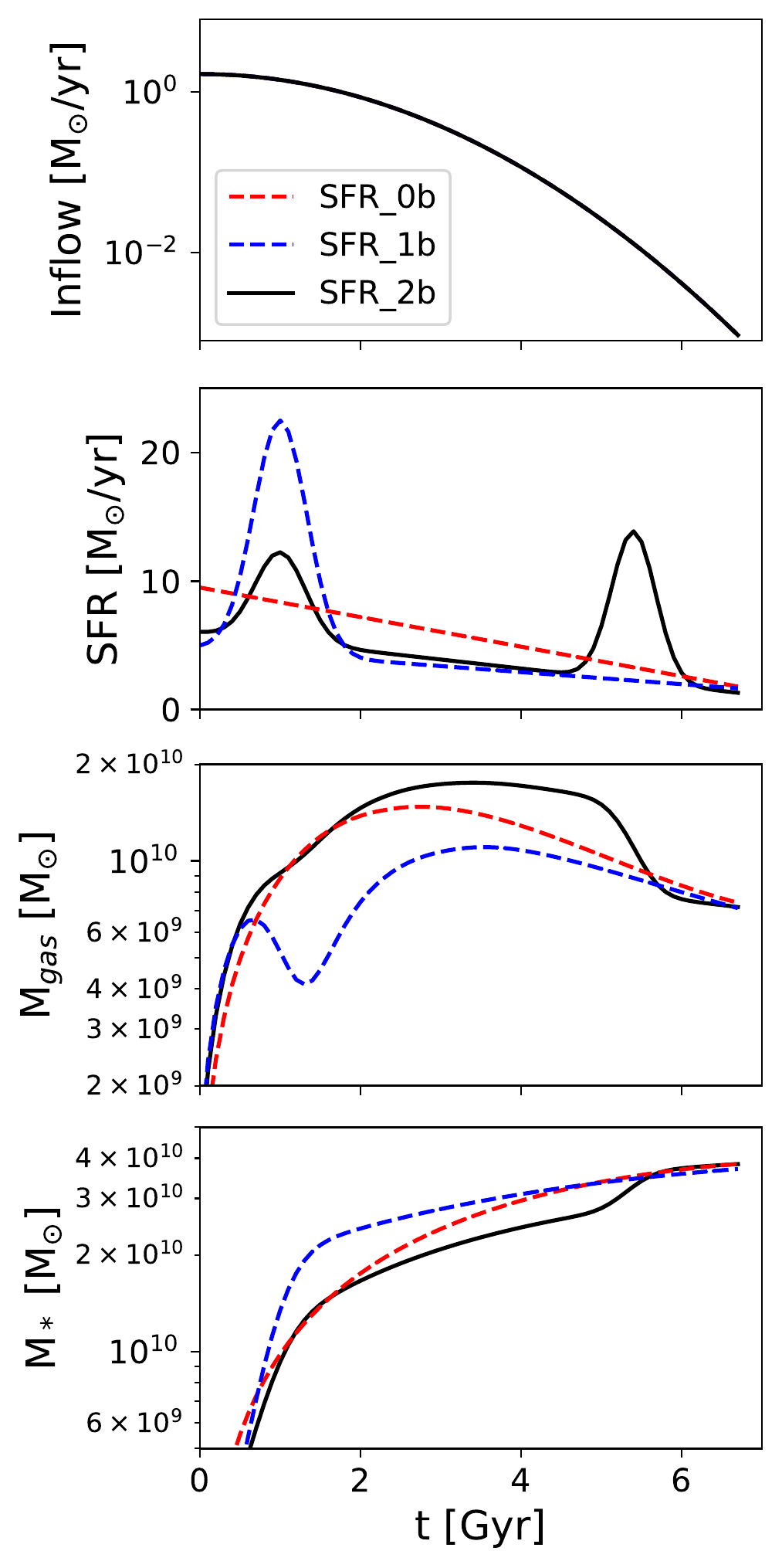}
    \caption{Gas inflow and SFHs in our main models. Top: Gas inflow history. Middle top: SFH. Middle bottom: Evolution of the total gas mass. Bottom: Stellar mass growth. In all panels solid black lines show the model that best reproduces the observed A(Li) evolution of thin-disk stars (SFR\_2b), dashed blue lines show the single star formation burst model (SFR\_1b), and the dashed red lines show the monotonically decreasing star formation model (SFR\_0b).}
    \label{Fig:ChemModICs}
\end{figure}

The inputs are the gas infall history (GIH), the SFH, and the initial A(Li) and [Fe/H]. We used a GIH and SFH that reproduce the observed stellar mass, gas mass, and star formation rate (SFR) in the MW thin disk at present: $\sim 3.5 $~\,$10^{10}\,\mathcal{M}_\odot$, $\sim 7 10^9\,\mathcal{M}_\odot$, and $\sim$1.65$\,\mathcal{M}_\odot$\,yr$^{-1}$, respectively \citep[][see also our Fig.~\ref{Fig:ChemModICs}]{BlandHawtorn2016}. Our three main models are SFR\_0b, SFR\_1b, and SFR\_2b, which only differ in the number of star formation bursts that the galaxy suffered during the thin-disk formation (see Fig.~\ref{Fig:ChemModICs}). In this work we use a GIH that resembles the gas inflow triggered by pristine gas cooling from a warm-hot circumgalactic medium (CGM) with a Navarro-Frenk-White (NFW) density profile \citep{NFW1996}. This assumption was motivated by the fact that the MW generated the current thin disk after developing a warm-hot CGM that inhibits gas accretion through cold flows \citep{DekelBirnboim2006,Behroozi2019,PostiHelmi2019}.\footnote{We obtained the cooling times used for the GIH computation from a grid of density-temperature single-zone CLOUDY models \citep{Ferland2017} of low-metallicity gas at $z=0$.} We assumed that the infalling gas has a primordial metallicity \citep[i.e. A(Li)=2.7, from ][]{Steigman2010}. In the current work we used a Chabrier initial mass function (IMF), but other options have also been tested and are available upon request: three-slope IMF - Kroupa93 \citep{Kroupa1993}, Chabrier03 \citep{Chabrier2003}, Salpeter \citep{Salpeter1955}, and a broken power law.\\

The model includes the following four  $^7$Li production mechanisms.  The first is SNeII: Only stars with masses between 8 and 30\,$\mathcal{M}_\odot$ release metals and gas into the ISM through SNII explosions. We consider that more massive stars fail to produce SNe. Following \citet{Stockinger2020}, 90\% of the progenitor's mass is returned to the ISM. Each SNII releases 5 $\times$10$^{-10}\,\mathcal{M}_\odot$ of lithium, in agreement with predictions from theoretical models \citep[e.g.][]{Nakamura2010,Kusakabe2019}. We did not include a model for lithium enrichment by pre-SNII winds. We obtain the number of SNeII in each timestep by obtaining the formation rate of SNII progenitors from the SFR and the IMF. We also compute the alpha and iron peak elements released by SNeII using the metal yields provided by \citet{Kobayashi2011}. 

The second is the AGB from LIMSs: Using the AGB metal yields presented by \citet{Ventura2013} for low-metallicity stars and by \citet{Ventura2020} for super-solar metallicity, we release into the ISM the lithium, alpha elements, and total gas mass produced by the stars in the AGB. We compute the number of stars in the AGB at each timestep by using the selected IMF and the SFH and by applying a time delay for the star's lifetime in the MS \citep{Argast2000}.

The third is novae: Each nova event ejects 7\,10$^{-4}\,\mathcal{M}_\odot$ of gas \citep{Shara2010}, including $\sim $2.5\,10$^{-10}\,\mathcal{M}_\odot$ of $^7$Li. These values agree with those obtained from observations of nova events  \citep[e.g.][]{Munari2008,Izzo2015} and from models for low-mass white dwarfs \citep[WDs; e.g.][]{Hernanz2005,Starrfield2020}. Also, a nova event injects $\sim 5 10^{-5}\,\mathcal{M}_\odot$ of carbon, $\sim 2 10^{-4}\,\mathcal{M}_\odot$ of nitrogen, and $\sim$ 5\,10$^{-4}\,\mathcal{M}_\odot$ of oxygen according to models by \citet{Hernanz2005} and \citet{Li2016} and recent observations by \cite{Munari2008}. We obtain the number of nova events per year by scaling the observed value ($\sim50$nova/yr in \citet{Shafter2017}) to the accumulated white dwarf formation history (AWDFH). We compute the AWDFH first by obtaining the formation rate of carbon-oxygen WD progenitors (stars with masses between $0.8$ and $8\,\mathcal{M}_\odot$) using the selected SFR and IMF. We then apply a delay to the WD formation that is equal to each star's lifetime in the MS \citep{Argast2000}. Later, we compute the time that each star spends in the recurrent nova period, before the SNIa occurs, following \citet{Greggio2005}. Finally, we subtract all the WDs that went through a SNIa event from the AWDFH.

The fourth is galactic cosmic rays (GCRs): The most complex process to account for in our GCE model is the production of lithium from GCR spallation. Empirically, the fraction of lithium produced by GCRs can only be obtained indirectly from measurements of the $^9$Be abundance as this ion is only produced in CRs \citep{Smiljanic2009,Grisoni2019}. Differently from \citet{Grisoni2019}, who used $^7$Li/$^9$Be=7.6, which is only valid for cold stars that depleted $^6$Li \citep{Molaro1997}, we use ($^6$Li+$^7$Li)/$^9$Be=13.1, which can be used for hotter stars. We then use the $^7$Li/$^6$Li=2 relation, obtained from theoretical works on CR spallation and alpha-alpha fusion in the ISM \citep{Meneguzzi1971,Fields2002}, to find a direct relation between $^9$Be and $^7$Li. Using Eq.~7 in \citet{Grisoni2019}, we finally obtain Eq.~\ref{eq:3}, which allows us to obtain the $^7$Li abundance produced by GCRs from the ISM metallicity. Metallicity evolution is computed self-consistently in our model, both for alpha and iron peaks, including all the aforementioned production mechanisms and the SNIa. The result reproduces the observed metallicity trend in our N-star sample well (see Sect.~\ref{sec:FeH_rotation_planets}) and is in good agreement with results from the HARPS-GTO/AMBRE-HARPS samples \citep{Minchev2018}. One of the largest sources of uncertainty when modelling the GCR lithium production is how the metal mixing time is accounted for. The mixing of metals in gaseous galactic disks is a complex process that is still under study \citep[e.g.][]{Yu2021,Rennehan2021}; it depends on many free parameters, for example the turbulence generated by SNe, the diffusion, and the shear by the disk differential rotation. So, in light of the results from both theoretical models and observations presented by \cite{Rennehan2021}, \cite{Marsakov2011}, and \cite{Hayden2017}, we decided to take a simple approach and apply a `mixing factor' that allows us to control how much gas has been polluted by SNe and stellar winds at each timestep. To determine its value, we needed to account for: (i) the low mixing efficiency of the ISM and IGM gas, as suggested by the wide range of thin-disk star metallicities in theoretical models and observations, and (ii) the dense gas that is shielded from CRs \citep[see a full discussion in][and references therein]{Ivlev2018}. Finally, we decided to use a conservative value of 50\%.

\begin{equation}
    \log (^7Li/H)=-9.982+1.24[Fe/H]
    \label{eq:3}
\end{equation}

We also included two $^7$Li destruction mechanisms that occur inside stars and, thus, do not affect the evolution of the ISM $^7$Li abundance. The first is the Eggenberger mechanism (PMS): The first $^7$Li depletion mechanism we implemented is the destruction of lithium in the PMS of Sun-like stars \citep{Eggenberger2012}. This mechanism depends on the star's rotation speed ($\Omega$) and the protoplanetary disk lifetime ($t_d$). In our best model we use $\Omega$=10\,$\Omega_{\odot}$, where the Sun's rotation is 27.5\,days \citep{Xie2017} and $t_d$=6\,Myr. 

The second is depletion in the MS: The lithium depletion during the MS is described well in \citet{Dumont2021}. We implemented results from their  x\_nu\_R1\_a\_t6.425 model, that is, the one that best fits observations. Results from this model are well described by the polynomial function we show in Eq.~\ref{eq:4}, where t is in units of Gyr. 

It should be noted that these mechanisms were calibrated to reproduce abundances of Sun-like stars (i.e. reaching solar metallicity at the Sun's age). In the approach used here, we did not account for changes derived from variations in the initial metallicity. We argue that as the  [Fe/H] of most thin-disk stars does not show strong variations from the solar value (see Fig.~\ref{fig:9}), we do not expect significant changes in our conclusions when including a metallicity dependence. However, this will need to be confirmed in further studies when metallicity-dependent models become available.

\begin{equation}
    A(Li)=2.904-0.565t+0.069t^2-0.008t^3
    \label{eq:4}
\end{equation}

\subsection{Modelling of the A(Li) thin-disk star evolution}

In this work we focus on the A(Li) evolution of thin-disk stars. The GCE model has been set accordingly: (i) It spans a range of 7\,Gyr, that is, the thin disk lifetime; (ii) we assume that the thin disk was generated from a single infall of pristine gas that peaked 7\,Gyr ago and decayed later exponentially; (iii) the total accreted gas mass is the one that results from the sum of the current thin disk stellar mass and its gas mass content; (iv) in our best model (SFR\_2b) we use a decreasing SFH, compatible with the cosmic star formation slowdown \citep{Madau2014}, plus a two-peaked distribution that was obtained by a direct summation of two Gaussians, one centred at 6\,Gyr (look-back time) and a second at 1.5\,Gyr. The two-peaked distribution was motivated by results from \citet{Mor2019} and \citet{RuizLara2020}.

\begin{figure}
    \centering
    \includegraphics[width=0.45\textwidth]{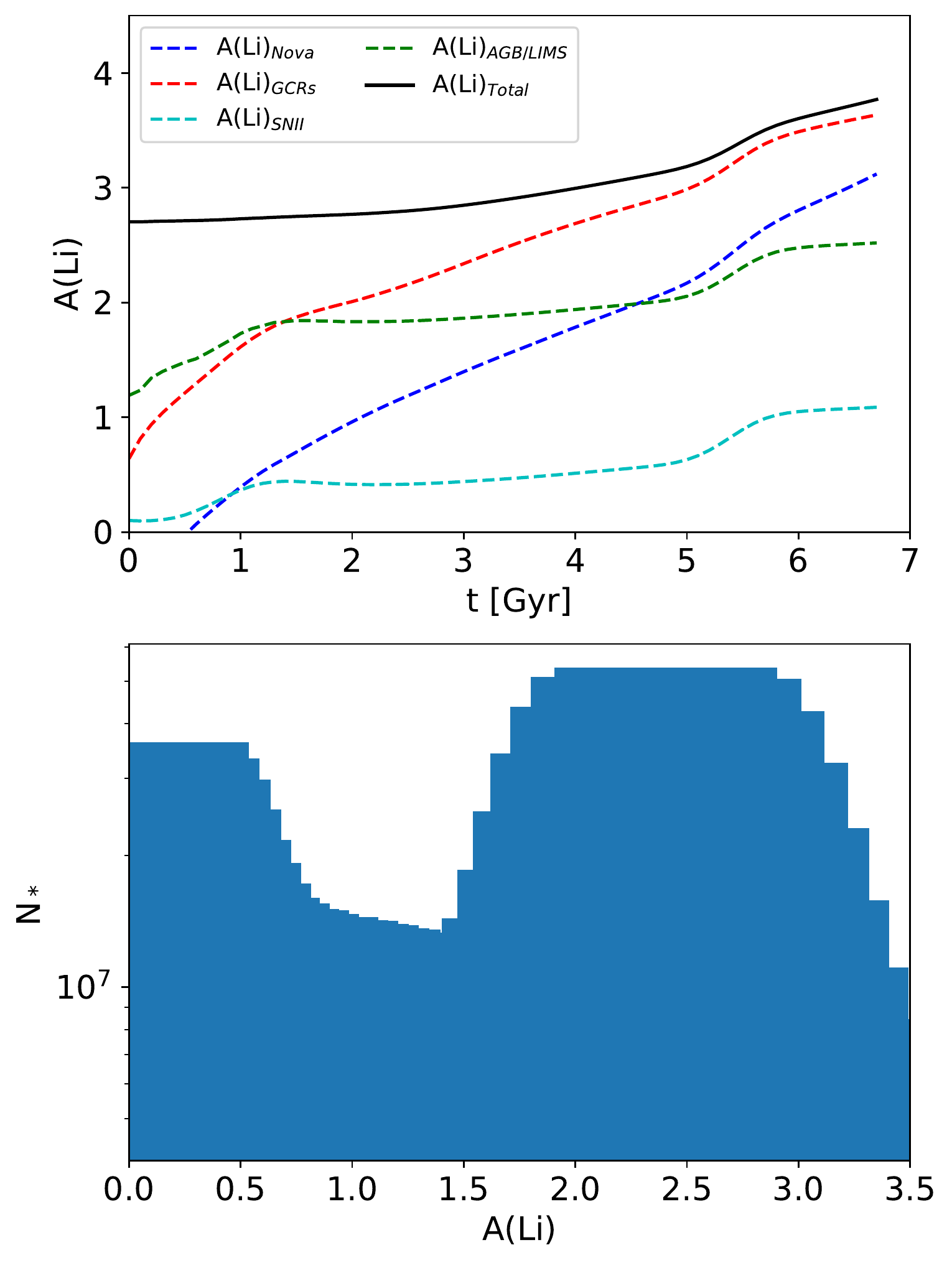}
    \caption{A(Li) evolution and its distribution in the FGK MS stars, obtained using our best GCE model. Top: A(Li) evolution in our best model (SFR\_b2) when not including the stellar depletion mechanisms described in Sect.~\ref{sec:Chem_model}. The solid black line shows the total A(Li) evolution. Dashed lines show the A(Li) computed using the lithium produced by each of the studied production mechanisms, without including the lithium mass at t=0; in red we show the contribution of GCRs, in blue the novae, in cyan the SNeII, and in green the AGB  and LIMSs. Bottom: Histogram of the total number of FGK MS stars per A(Li) bin generated by our best GCE model when including all aforementioned production  and depletion mechanisms. We obtained the bin size of all histograms in this work by applying the Freedman-Diaconis rule \citep{Diaconis2004}}
    \label{fig:ALi_evol}
\end{figure}

In Fig.~\ref{fig:ALi_evol} we show the 7\,Gyr A(Li) evolution obtained using our best model, SFR\_2b. In the top panel we show the ISM A(Li) evolution when the stellar depletion mechanisms are not included. In the bottom panel we show the resulting 7\,Gyr evolution of stellar A(Li) when the two stellar depletion mechanisms described in Sect.~\ref{sec:Chem_model} are included. We observe that GCR spallation is the dominant mechanism in a system that suffered a former quenching process (e.g. low SFR before t=0 in our model). After 5$-$6\,Gyr of continuous star formation, nova contribution becomes more important and could potentially be the dominant mechanism of $^7$Li production at later ages, in overall agreement with recent works by \citet{Grisoni2019} and \citet{Dumont2021}. It is important to note here that the contribution of the GCR spallation is sensitive to the efficiency of gas mixing in the IMF and that this process is still poorly understood (see Sect.~\ref{sec:Chem_model}). $^7$Li production by SNeII, AGB stars, and LIMSs are subdominant mechanisms throughout the simulation.\\

Our GCE model will be updated in future works to include a model for gas mixing by SN turbulence and GCR emission, dependences on the galactocentric distance, and radial migration. These modifications will allow us to expand the analysis from global to local properties of the MW thin disk.

\section{The A(Li) distribution: Two stellar populations connected by an isthmus} \label{sec:isthmus}

From the first works that studied the distribution of A(Li) in stars, as a function of their surface temperature ($T_{\rm eff}$), a notorious rarefied field of stars with A(Li) $\sim 1.5$ appeared \citep[see for instance][]{Chen2001}. More recently, new surveys containing larger stellar catalogues  \citep{Ramirez2012,BensbyLind2018,AguileraGomez2018} have confirmed the existence of this stellar population. These works also pointed towards the presence of an under-abundance of stars in the same A(Li) region when using stars with planets only. This apparent gap for stars that host planets can be seen in Fig. 9a of \citet[][hereafter RA12]{Ramirez2012}. Lately, with the advent of new and larger surveys that double the number of stars compared to RA12 (e.g. AG18), this apparent gap of stars with planets tended to disappear (see Fig. 2 in AG18).

\begin{figure}
    \centering
    \includegraphics[width=0.49\textwidth]{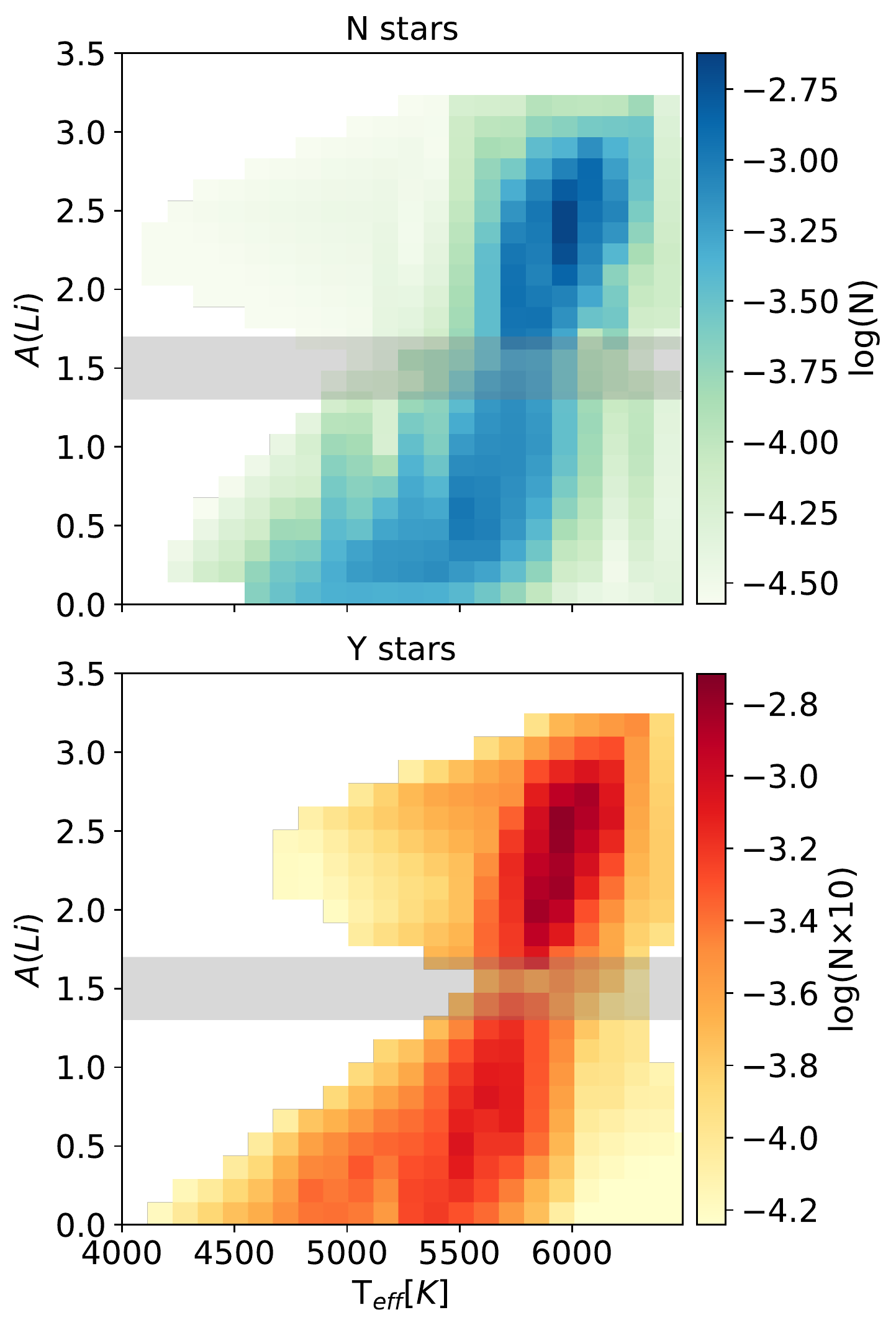}
    \caption{A(Li) vs. $T_{\rm eff}$ for stars without known planets (N, {\em top}) and those that host confirmed planets (Y, {\em bottom}). The grey shaded region corresponds to the transition region we called the isthmus. The colour bar shows the logarithm of the number of stars per A(Li)-$T_{\rm eff}$ bin, normalised to unity for N stars and the same value multiplied by ten for the Y stars. 
}
    \label{fig:ALi_Teff}
\end{figure}

\begin{figure*}
    \centering
    \includegraphics[width=0.95\textwidth]{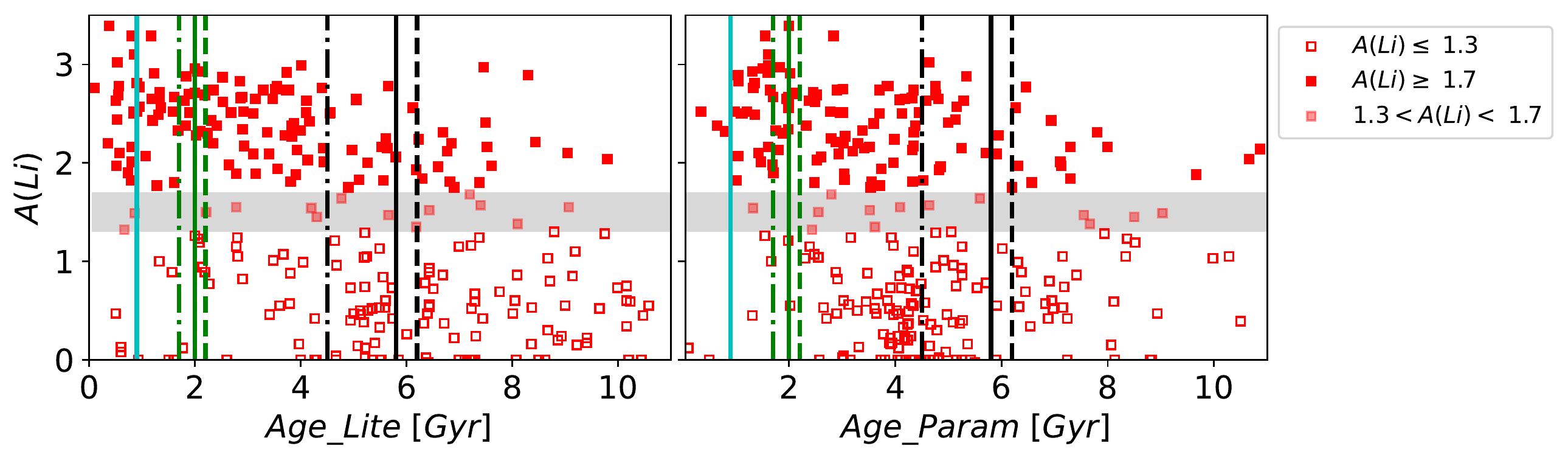}
    \caption{A(Li) vs. the two age scales used in this work (see Sect.~\ref{sec:observations}) for stars with planets (Y). {\em Left panel}: Age from the literature ($Age\_Lite$). {\em Right panel}: Age from isochrones ($Age\_Param$). The vertical lines represent the main galactic events known to happen in that temporal range (Sagittarius dwarf pericentres and the star formation burst presented in \citealt{Mor2019}). Consistent with previous figures, the grey shaded region corresponds to the transition region we called the isthmus.}
    \label{fig:ALi_Age_2}
\end{figure*}

In this work we used a much larger sample of stars compared to previous works, such as AG18 (see Sect.~\ref{sec:observations}). In addition, we analysed the A(Li) versus $T_{\rm eff}$ distribution of both stars with (Y) and without (N) planets (see Fig.~\ref{fig:ALi_Teff}). We confirm that the under-density of stars near A(Li)$\sim1.5$ is a singular feature in the general distribution of the lithium abundance visible in both N and Y populations, and we call this the isthmus\footnote{A narrow piece of land with water on each side that joins two larger areas of land \citep{Britannica2006}.}. In Fig.~\ref{fig:ALi_Age_2} we show the A(Li) distribution of Y stars as a function of the two age scales described in Sect.~\ref{sec:observations}. We observe that the isthmus may also be a transition between two different stellar age groups.

In our effort to understand the origin and properties of the isthmus,
we assumed the hypothesis that it is in fact a transit region between
the two groups, the A(Li)-rich and the A(Li)-poor. If this hypothesis
is correct, we should find a mechanism that is able to produce a
significant increase in the $^7$Li depletion, that is, the transition.
We started by studying the relation between stellar activity and the observed changes in the A(Li) distribution, following results obtained by \citet{Li2014}. In that paper the authors presented a model where age and stellar activity are correlated and which could explain the observed age-A(Li) relation. This model includes a Tayler-Spruit dynamo field, which produces an extra-mixing process that is directly related with stellar activity and A(Li) depletion. However, we noticed that if this were the only physical mechanism that shaped the A(Li) distribution, the age-A(Li) correlation would be much tighter and the presence of two epochs in which the slope changes and becomes very steep ($\sim$2\,Gyr, and $\sim$4\,Gyr in Fig.~\ref{fig:ALi_Age_2}) would not be explained. In Sect.~\ref{sec:galactic} we describe the physical mechanisms that can reproduce the observed distribution. \smallskip

In the previous paragraphs we presented a qualitative definition of what we call the isthmus. Here we present the quantitative definition that we use all through this paper, obtained by choosing a region centred on the A(Li) `gap' (A(Li)$\sim1.5$) and with a width equal to the maximum error on the A(Li) (i.e. 0.1). In Figs.~\ref{fig:ALi_Teff} and \ref{fig:ALi_Age_2} we show how by using this definition ($1.3<$A(Li)$<1.7$) we are properly selecting the isthmus region in both N and Y stars (see the grey shaded region). Looking at the A(Li)$-T_{\rm eff}$ distribution, we found that inside the isthmus region there are 15 Y stars and 106 N stars. This is a number large enough to confirm that all regions have at least some number of both Y and N stars; as such, the isthmus region is not empty but just under-dense. In the next section we study the origin of this under-density and its possible relation with the MW thin disk formation history.

\
Next we wanted to confirm that the assumption of A(Li) being a bi-modal distribution is statistically significant. With this purpose, we applied the Kolmogorov-Smirnov (K-S) test to the observed N-star and Y-star A(Li) distributions. Our null hypothesis is that the A(Li) distributions are bi-Gaussian (see Appendix~\ref{app:KS} for further details). For the N and Y samples, we obtained a p value of 0.11 and 0.39, respectively, which is well above the standard significance level that allows us to accept the null hypothesis. In Table~\ref{tab:centroids} we show that the N stars are well described by a combination of a A(Li)-rich Gaussian distribution of stars centred at 2.25 with a standard deviation of 0.5, and another A(Li)-poor distribution of stars centred at 0.5 with a standard deviation of 0.35. In Table~\ref{tab:centroids} we show that Y stars, like N stars, are well described by two Gaussian distributions: the A(Li)-rich population centred at 2.4, with a standard deviation of 0.3, and the A(Li)-poor centred at 1.1, with a standard deviation of 1.15. Our definition of the isthmus region fits well within the space left between the two populations (grey shaded region in Figs.~\ref{fig:ALi_Teff} and \ref{fig:ALi_Age_2}). \\ \smallskip

\begin{table*}
\centering
\setlength{\tabcolsep}{3.5pt}
\renewcommand{\arraystretch}{1.65}
\caption{Mean A(Li) and age for the A(Li)-rich and A(Li)-poor populations.} 
\label{tab:centroids}
\small
\begin{tabular}{ c c c| c c c}
\hline
\multicolumn{6}{c}{\bf Y stars} \\
\hline
\multicolumn{3}{c|}{A(Li) $\ge 1.7$} & \multicolumn{3}{c}{A(Li) $\leq 1.3$} \\
\hline
A(Li)   & $Age\_Param$ & $Age\_Lite$  & A(Li) &   $Age\_Param$ & $Age\_Lite$\\
 &   \footnotesize{[Gyr]} & \footnotesize{[Gyr]} &    &    \footnotesize{[Gyr]} & \footnotesize{[Gyr]}\\
\hline      
$2.40\pm0.02$  & $3.1\pm0.3$ & $2.9\pm0.3$ & 
$1.05\pm0.05$  & $4.70\pm0.16$ &  $4.8\pm0.2$  
 \\
\hline
\hline
 \multicolumn{6}{c}{\bf N stars} \\
\hline
\multicolumn{3}{c|}{A(Li) $\ge 1.7$} & \multicolumn{3}{c}{A(Li) $\leq 1.3$} \\
\hline
A(Li)   & $Age\_Param$ & $Age\_Lite$ &   A(Li) &   $Age\_Param$ & $Age\_Lite$\\
 &   \footnotesize{[Gyr]} &  \footnotesize{[Gyr]} &  &   \footnotesize{[Gyr]}&   \footnotesize{[Gyr]}\\
\hline      
$2.37\pm0.01$ &  $4.34\pm0.12$ & $4.48\pm0.13$ & 
$0.47\pm0.01$ &  $5.28\pm0.14$ & $5.57\pm0.16$   \\

\hline
\end{tabular}
    \tablefoot{In this table we show results for the two age scales used in this work and their standard error. We show results for both the A(Li)-rich and A(Li)-poor populations. We also include values for the Y and N samples. The values listed in this table have been computed in the two A(Li) regions defined by the K-S test presented in Appendix~\ref{app:KS}.}
\end{table*}

\section{The Milky Way star formation history from changes in the A(Li) distribution} \label{sec:galactic}

The chemical evolution of our Galaxy can be studied from the metallicity observed in stellar atmospheres \citep[e.g.][]{McWilliam1997}. Variations in metal abundances in stars of different ages can be a good tracer of the large-scale events that shaped the galactic SFH \citep{Yong2008,Lin2020}. It is expected that a bursty star formation and strong low-metallicity gas inflows from the CGM or the IGM leave characteristic imprints on the observed stellar chemistry. In this section we focus on the properties of the stellar A(Li) distribution of thin-disk stars and, in particular, on its dependence on stellar age. Using our GCE model (see Sect.~\ref{sec:Chem_model}), we discuss the role that Galactic-scale events played in shaping the lithium abundance distribution of MW thin-disk stars.

\subsection{A(Li) evolution and the galactic events: The origin of the isthmus}\label{sec:Li_age}

As discussed in the previous sections, A(Li) evolution in stars depends on several mechanisms: (i) the initial abundance of the star-forming molecular clouds --  primordial gas accretion, pollution by stellar feedback, $^7$Li production from CRs, or depletion in stellar accretion disks by the Eggenberger mechanism; (ii) $^7$Li depletion in burning layers of MS stars (see LA21); (iii) $^7$Li production by the Cameron-Fowler mechanism -- production in the atmospheres of the LIMS red giants \citep[][and references therein]{CameronFowler1971};
and (iv) enhancement by planet engulfment that induces a redistribution of the lithium in the convective envelope of the host star \citep{AguileraGomez2016,Stephan2020,SoaresFurtado2020}.

These mechanisms strongly depend on the stellar age and metallicity, and, therefore, the galactic A(Li) evolution is strongly coupled with the SFH and gas accretion history \citep[GAH; see Sect.~\ref{sec:Chem_model} and][]{LivioPringle2003,SanchezAlmeida2018,CescuttiMolaro2019}. We argue that the $^7$Li evolution can be used to identify and characterise Galactic-scale events that affected the SFH and/or the GAH \citep[e.g.][]{SanchezAlmeida2014,SanchezAlmeida2014b}. In particular, while gas inflows have a small impact on the galactic SFR, strong interactions with neighbour galaxies are a good example of events that strongly affect their SFH. We discussed in previous sections how these interactions induce a sudden increase in the host galaxy's SFR, a process that is commonly known as a star formation burst. This burst consumes a fraction of the gas available for star formation and quickly pollutes the remaining gas with metals from core-collapse SNeII first, and later from AGB winds, novae, SNeIa, and neutron stars mergers \citep{Lin2020}. Stars formed during the `pollution period' or later show clear signs of the latest star formation burst in their chemistry. In particular, the newly formed stars will exhibit a similar initial A(Li) that is different from the one in the former generation of stars. Additionally, all stars born in the same star formation burst will show a synchronised starting point in their lithium depletion processes. First, in most stars $^7$Li will suffer a strong depletion process in the PMS through the Eggenberger mechanism. Later, in the bulk of FGK-dwarf stars, $^7$Li will be slowly, but constantly, destroyed during their MS. In this last process stars will transit from the A(Li)-rich region to the A(Li)-poor region, across the A(Li)-age diagram. Qualitatively, an old star formation burst will be observed as a A(Li)-poor dense clump in the A(Li)-age space, while a recent event will be seen as a clump in the A(Li)-rich region. A constant star formation process spanning several gigayears will be observed as a continuous path connecting the old A(Li)-poor and the young A(Li)-rich populations (see Fig.~\ref{fig:ALi_evol}, bottom panel); this is what we call the isthmus.

\subsection{Two recent star formation events hidden in the A(Li) distribution}\label{sec:sf_bursts}

\begin{figure*}
    \centering
    \includegraphics[width=0.45\textwidth]{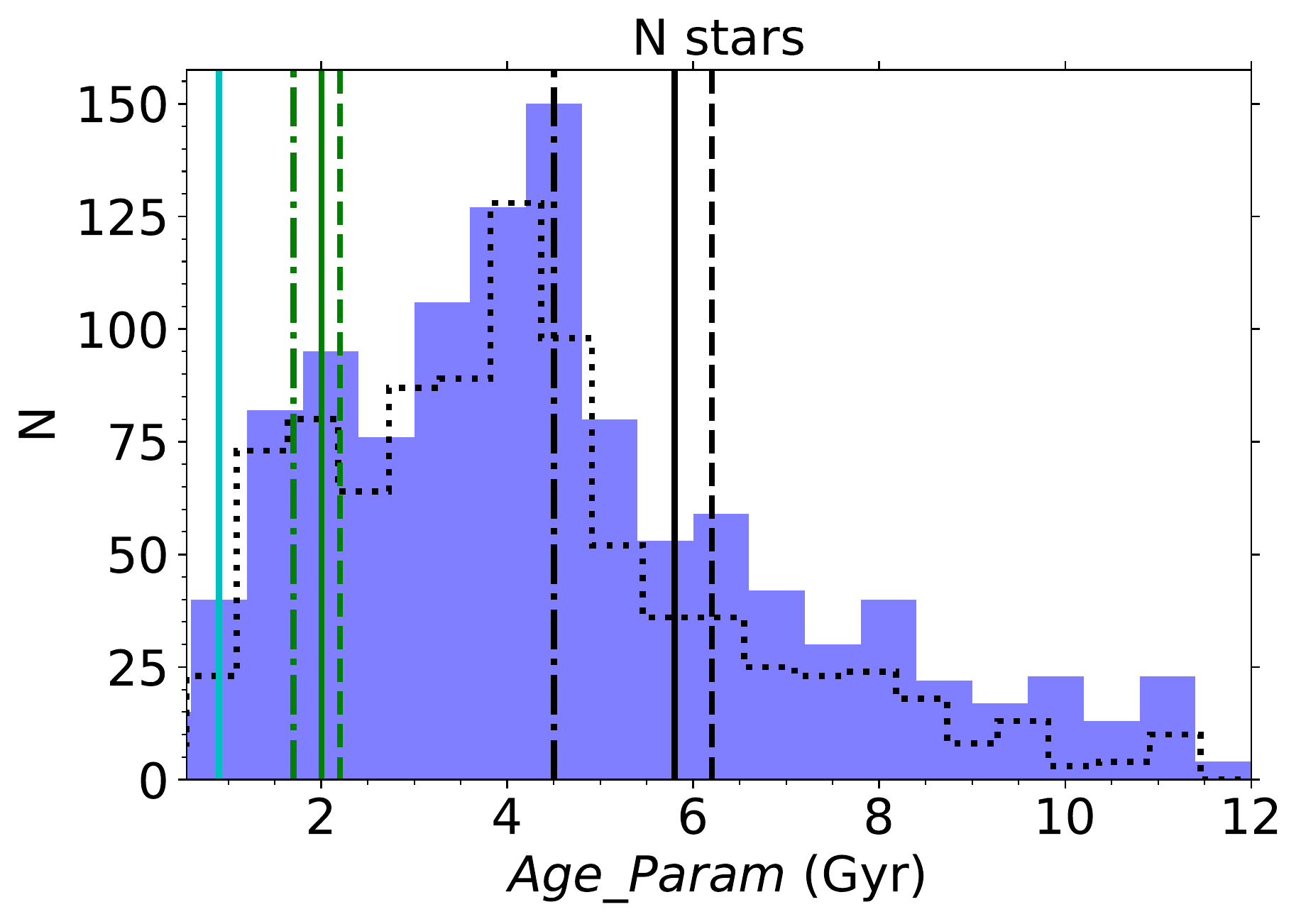}
    \includegraphics[width=0.45\textwidth]{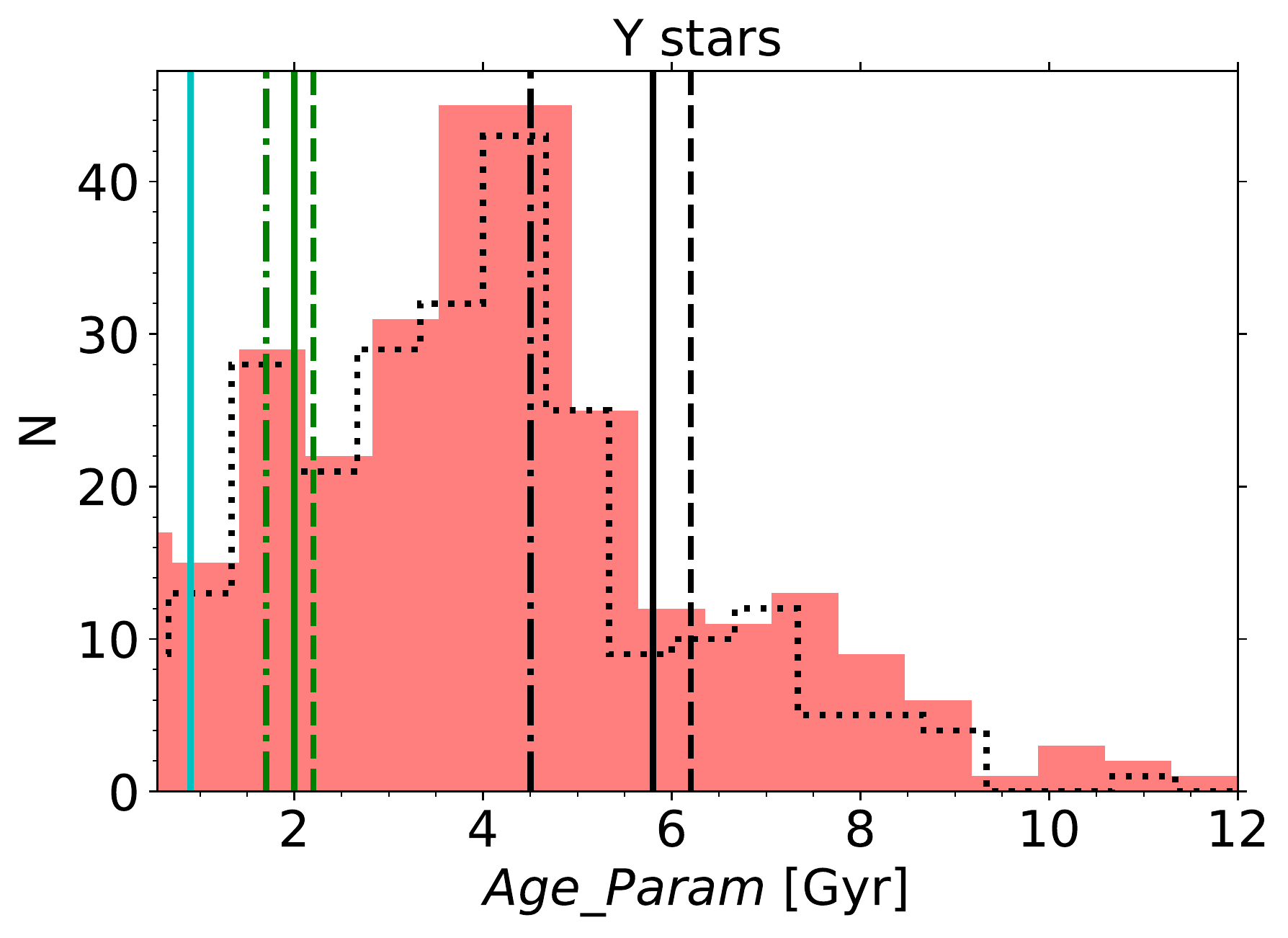}    \caption{$Age\_Param$ histogram for stars with ({\em right}) and without ({\em left}) planets. We show the thin disk contribution to the total with a black dotted histogram. The bin width has been obtained by applying the Freeman-Diaconis rule. Vertical lines show the start (dashed), peak (solid), and end (dot-dashed) of MW star formation bursts presented in \citet{RuizLara2020}; the lines in black show the strong star formation burst detected in \citet{Mor2019}, in green the last Sagittarius dwarf galaxy pericentre, and in cyan a very recent star formation burst also detected in the local stellar kinematics \citep{Antoja2018}.}
    \label{fig:7}
\end{figure*}

Using the GCE model presented in Sect.~\ref{sec:Chem_model}, we can set a link between our A(Li) data from thin-disk FGK MS stars (Sect.~\ref{sec:observations}) and the well-known star formation events in the MW. First, we analysed the age distribution of stars in our sample and looked for imprints of the star formation events recently discovered by \citet{Mor2019} and \citet{RuizLara2020}. In Fig.~\ref{fig:7} we show histograms of the number of stars without planets, N stars (left panel), and with planets, Y stars (right panel), as a function of $Age\_Param$. We show the data in a number of bins obtained by applying the Freedman-Diaconis rule \citep{Diaconis2004}. The bin width is always larger than $\sim$0.6\,Gyr, which is of the order of half the average age uncertainties of the young stars in our sample (see LA21 for details on the age uncertainties). In order to ensure our results are independent from the age determination technique, we compared values presented here with the most recent values in the literature (i.e. $Age\_Lite$; see Appendix~\ref{app:Ages}). To better understand the impact of the Galactic-scale events on the different galactic components (i.e. both thin- and thick-disk), we also show the contribution of the thin-disk stars to the total number of stars as a dashed histogram curve (see the decomposition strategy in Sect.
~\ref{sec:data_kin}). Vertical lines show the start (dashed), peak (solid), and end (dot-dashed) of the star formation events observed in {\em Gaia} DR2 by \citet{RuizLara2020}. These events almost coincide with those reported by \citet{Mor2019} and \citet{Isern2019}. In both distributions, N (left) and Y (right), the main star count peaks and the two main galactic star formation events reported in the literature (black and green vertical lines) almost overlap. Apart from star formation peaks, we also detected a clear star formation quenching following such events, something that we expect from theory of galaxy formation \citep[e.g.][]{Maltby2018}. It is important to note here that these results are independent of the age scale that we used in our analysis ($Age\_Param$ or $Age\_Lite$). The coincidence between results from previous works and our data is not perfect due to uncertainties in both modelling \citep[see][]{Mor2019,Isern2019,RuizLara2020} and age determination techniques (average uncertainties in our sample are of $\sim2$\,Gyr), which introduce a certain degree of dispersion in our results. The more evident discrepancy is on the position of the peak of the main star formation event at $\sim$4$-$6\,Gyr. For this event our star count distribution peaks when the reported star formation event reaches its final phase (dot-dashed line). Although uncertainties in the age determination make it difficult to obtain strong constraints on the SFH shape from the A(Li) distribution, we argue that the main conclusions that we reach here are not affected as they do not depend on the age scale.\\ 

\begin{figure*}
    \centering
    \includegraphics[width=0.45\textwidth]{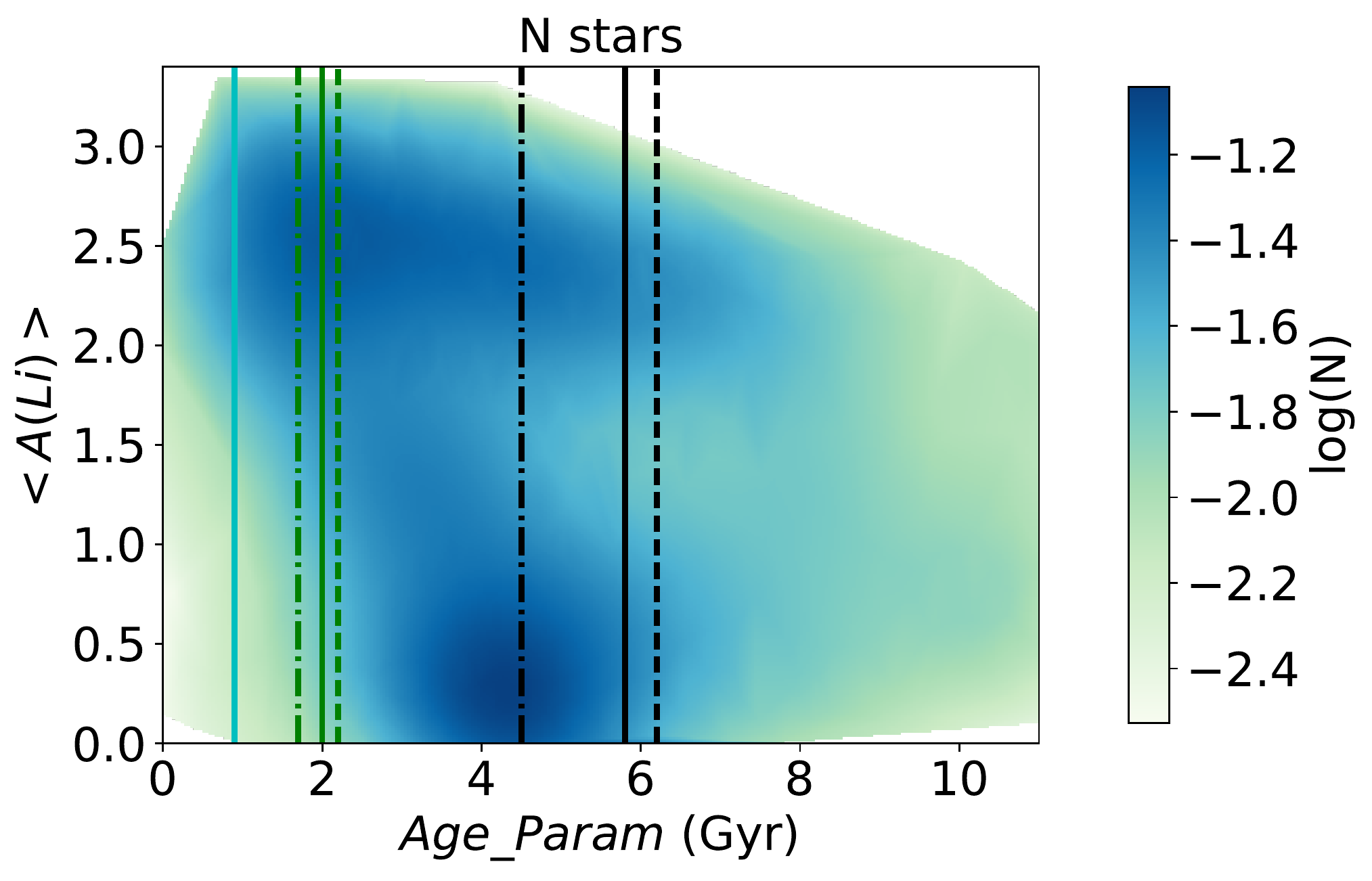}
    \includegraphics[width=0.45\textwidth]{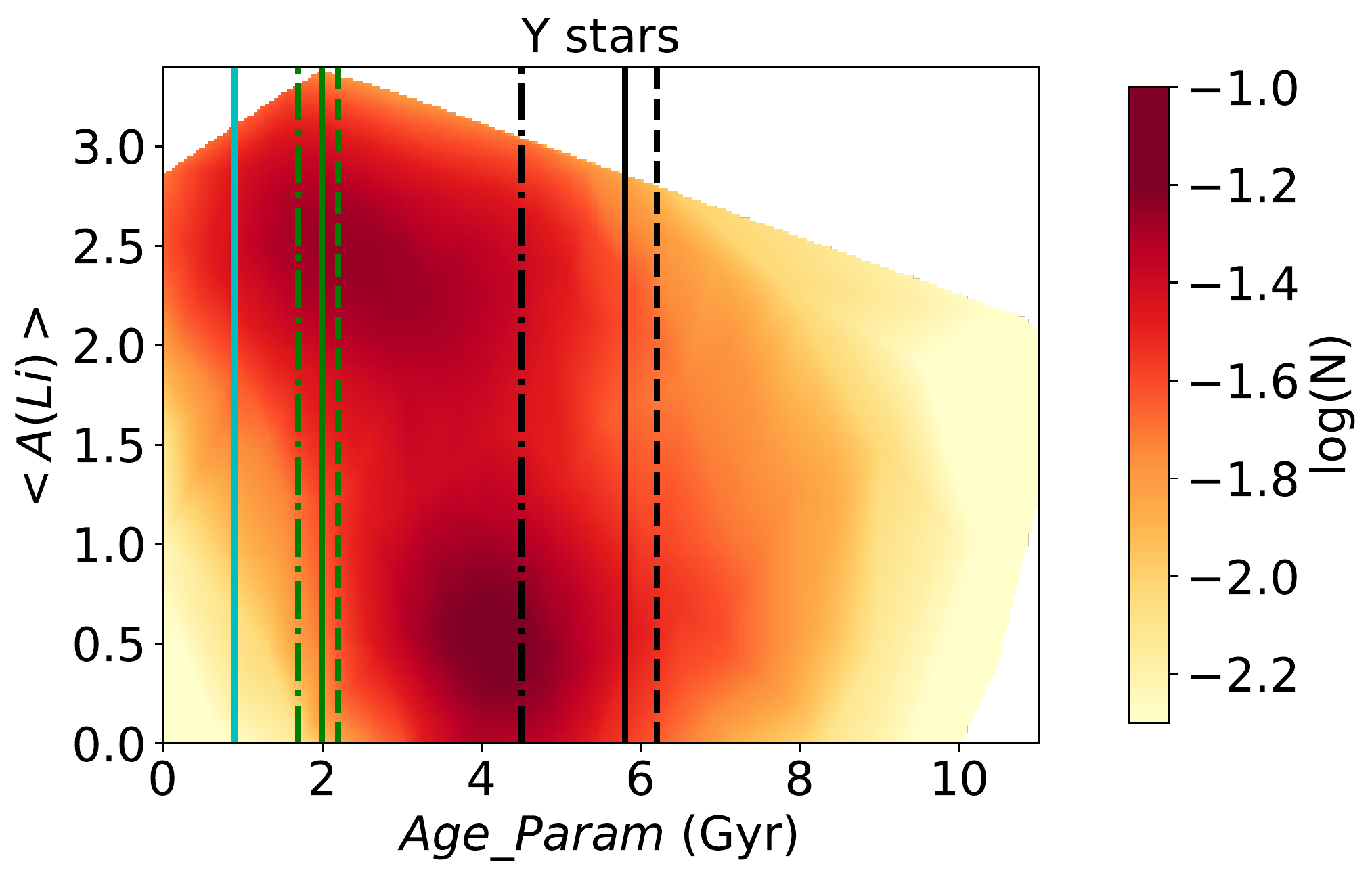}
    \caption{Mean A(Li) computed in 0.1\,dex bins vs. mean $Age\_Param$ in 0.6\,Gyr bins, for stars without known planets (N, {\em left}) and those that host confirmed planets (Y, {\em right}). Vertical lines show the last star formation burst in \citet{RuizLara2020} (line colours and styles as in Fig.~\ref{fig:7}). Colour bars show the logarithm of the number of stars per A(Li)-age bin, normalised to unity. Equivalent plots with only thin- and/or thick-disk stars only are available upon request to the authors.}
    \label{fig:8}
\end{figure*}

After confirming that in our sample we detect the two star formation events that generated the thin disk according to \citet{Mor2019}, \citet{Isern2019}, and \citet{RuizLara2020}, in Figs.~\ref{fig:8} and \ref{fig:9} we analyse the chemical evolution of N and Y stars in our sample (we also show the corresponding figures when using $Age\_Lite$ instead of $Age\_Param$ in Appendix~\ref{app:Ages}). A first look at the A(Li) evolution (Fig.~\ref{fig:8} using $Age\_Param$, and $Age\_Lite$ in Appendix~\ref{app:Ages}) confirms our qualitative analysis from Sect.~\ref{sec:Li_age}: Two A(Li) stellar populations exist -- one young and A(Li)-rich and one old and A(Li)-poor, coinciding with the recently discovered star formation events -- as does a low density region that connects them via what we named the isthmus. The A(Li)-poor population ranges from near-solar \citep[$\sim$1.1\,dex in][]{Carlos2019} to 0.0\,dex, with the region of  the isthmus ranging from $\sim$1.3\,dex to $\sim$1.7\,dex. It is notable that this population emerges just after the star formation burst initiated $\sim$6.5\,Gyr ago and spans a time interval that includes the epoch of the Sun formation ($\sim$4.5\,Gyr ago). The A(Li)-poor population, both in Y and N stars, disappears at ages younger than $\sim$4\,Gyr in favour of the transition population of the isthmus. The A(Li)-rich population takes values between 1.7 (slightly below the Spite plateau) and 3.2 (the observed value for newborn thin-disk stars). Differently from the A(Li)-poor population, the evolution of N and Y A(Li)-rich stars differs: The A(Li)-rich N population is detected long before the aforementioned star formation event, while the A(Li)-rich Y population seems to appear with the star formation burst. It is in the A(Li)-rich N population that we detect a third population that is A(Li)-rich and spans a wide range of ages. Our preliminary analysis indicates that stars in this third population have a particularly low rotation velocity. In a future paper we will present a study of how planet engulfment events can affect stellar rotation and replenish the stellar atmosphere with new $^7$Li, as suggested by \citet{AguileraGomez2016}, \citet{Stephan2020}, and \citet{SoaresFurtado2020}. The fact that in the left panel of Fig.~\ref{fig:8} the increase in the mean A(Li) in the A(Li)-rich N population changes its slope just after the star formation event ($\sim$4.5\,Gyr), from a slow increase to saturation at about A(Li)$\sim$3.2, may support the hypothesis of planet engulfments being the drivers of the formation of the third population. This situation could indicate a reset and restart of the A(Li) depletion mechanisms of stars that underwent this sudden A(Li) enhancement.\\

\begin{figure}
    \centering
    \includegraphics[width=0.45\textwidth]{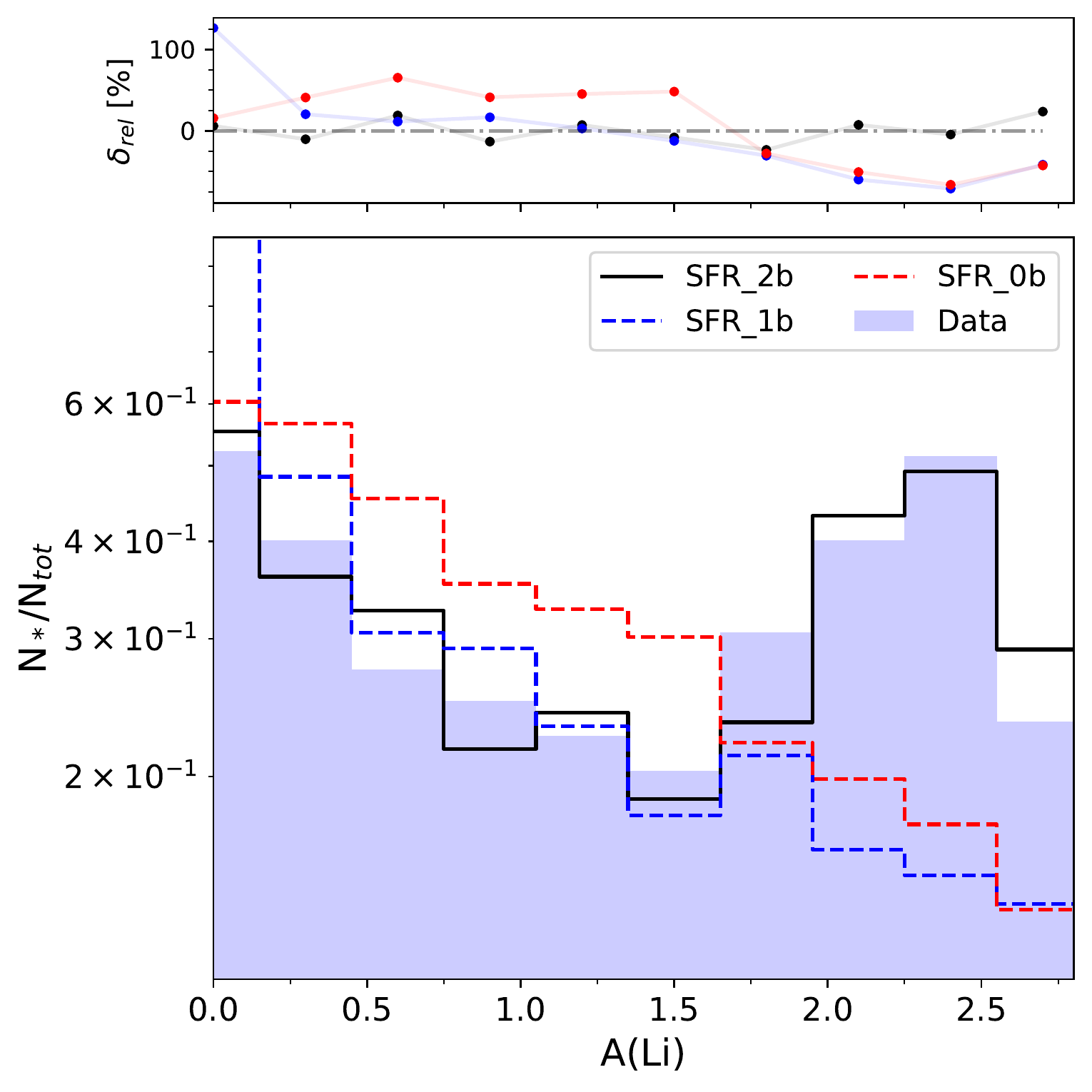}    \caption{Normalised A(Li) histogram of all stars in our data sample (soft blue colour bars) vs. predictions by our GCE models (solid and dashed lines), and relative errors (top panel). Dashed red lines show the results from our SFR\_0b model, dashed blue lines from SFR\_1b, and solid black lines from the SFR\_2b model, which is the one that best reproduces the observed A(Li) distribution (see Sect.~\ref{sec:Chem_model} for more information on the models). Bin numbers in all histograms have been obtained by applying the Freedman-Diaconis rule \citep{Diaconis2004}.}
    \label{fig:Data_vs_model}
\end{figure}

Finally, we confirm our hypothesis on the origin of the bi-modal A(Li) distribution, and of the isthmus, by comparing results from our GCE model with data. In Fig.~\ref{fig:Data_vs_model} we show the normalised A(Li) distribution of all the stars in our sample (blue histogram) compared with the result from three models of GCE: (i) imposing a gently declining SFH (SFR\_0b model, dashed red lines); (ii) a gently declining SFH plus a single star formation burst at a $\sim$6\,Gyr look-back time (SFR\_1b model, dashed blue lines); and (iii) our best model (SFR\_2b, black-solid lines), which contains a gently declining SFR plus two star formation bursts at $\sim$6\,Gyr and $\sim$1.5\,Gyr look-back times, the latter being the most impulsive. In our best model the first star formation peak is only 68\% as intense as the latter and lasts 25\% longer. The SFR\_2b model is the one that best reproduces the observed A(Li) distribution while recovering the presently observed SFR and the thin-disk stellar and HI mass \citep{Mor2019,Isern2019,RuizLara2020}. This result is robust as it only depends on the observed A(Li) distribution. Possible biases from the determination of the stellar ages do not have an impact on our conclusions. Figure~\ref{fig:Data_vs_model} clearly shows that only a two-peaked SFH can explain the observed bi-modal A(Li) distribution when accounting for most known $^7$Li production and depletion mechanisms (solid black line). It is important to remark that we cannot quantify the exact intensity and length of the star formation bursts from our data and models, but we can only give a qualitative analysis. This is a consequence of: (i) the large age determination uncertainties still present in current data; (ii) the fact that the data we use is restricted to the solar neigbourhood and thus do not represent the entire Galaxy; and finally (iii) the fact that our model still misses some $^7$Li production and destruction processes that are currently under study (e.g. the effect of planet engulfment).  In the future, new data from large surveys such as Galah \citep{Buder2021} combined with better determinations of A(Li) in $^7$Li-poor low-luminosity cold dwarf stars will result in large, unbiased samples of MS stars. These new catalogues will allow us to better understand the many channels of $^7$Li production and destruction that occur inside dwarf stars. Also, future catalogues that include precise age determinations will allow us to better constrain the SFH of the MW, also by analysing its A(Li) distribution.

\subsection{[Fe/H], stellar rotation, and planets}\label{sec:FeH_rotation_planets}

In this last section we discuss how [Fe/H] evolution, stellar rotation, and the presence of planets can affect the A(Li) distribution. This because we cannot omit the global [Fe/H] evolution or processes linked to stellar evolution, such as variations in its rotation and the presence of planets, when building a global picture of the galactic A(Li) evolution. However, it is not within the scope of this paper to deeply analyse the role they play; here we only give a brief summary of the results that will be presented in future papers of our collaboration.\\

\begin{figure*}
    \centering
    \includegraphics[width=0.45\textwidth]{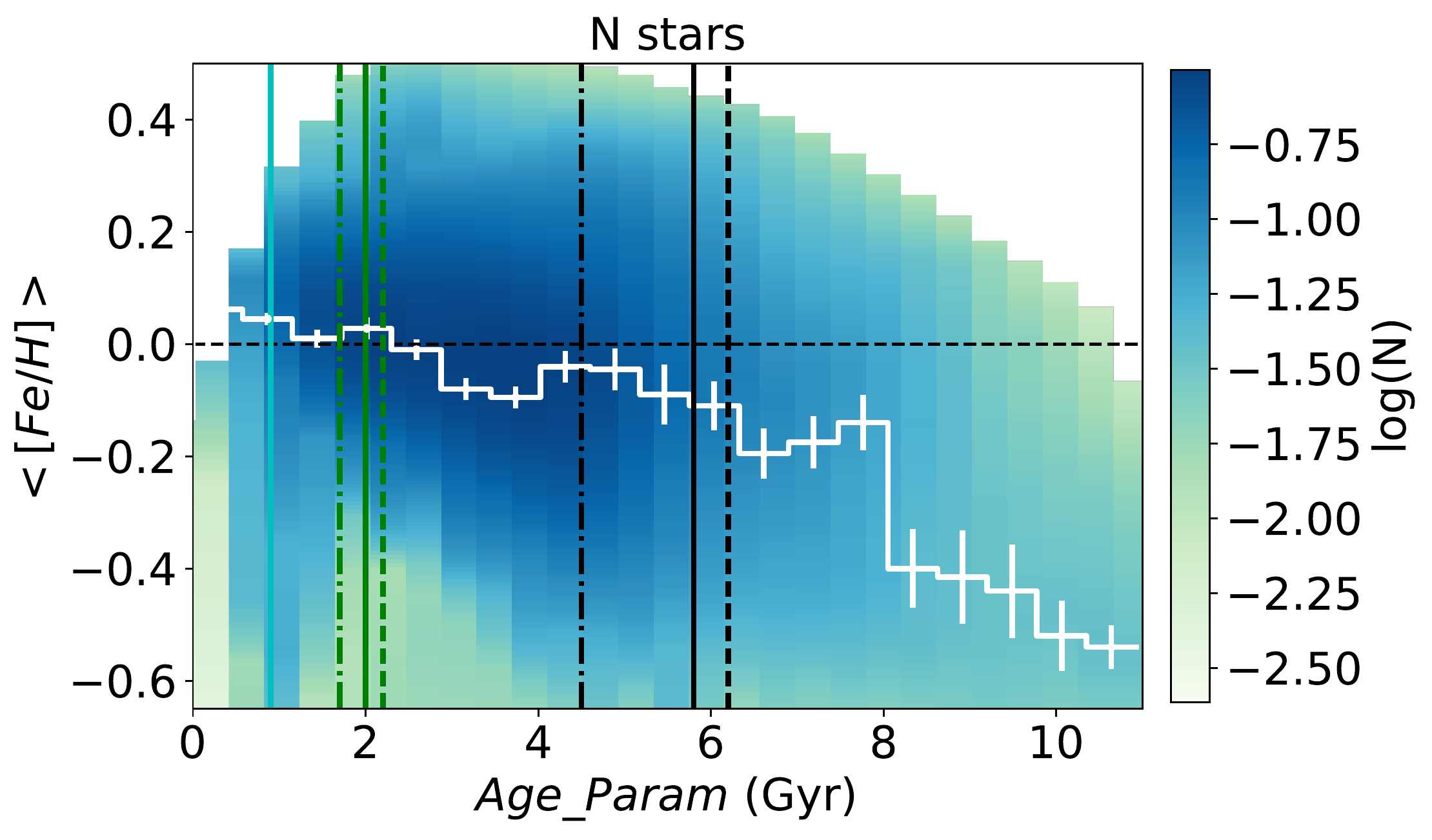}
    \includegraphics[width=0.45\textwidth]{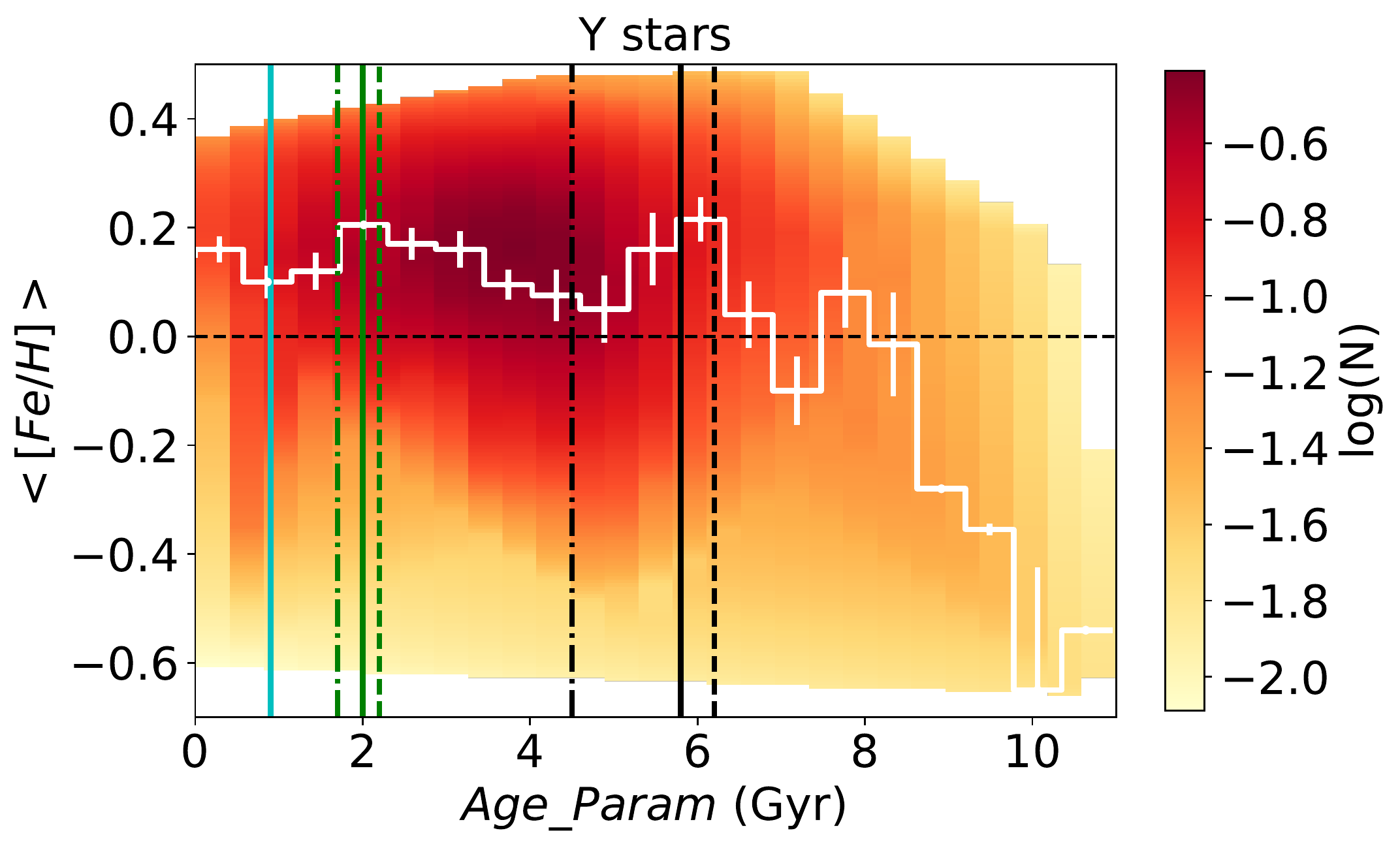}
    \caption{[Fe/H] vs. $Age\_Param$ density plot, for stars without known planets (N, {\em left}) and those that host confirmed planets (Y, {\em right}). White solid lines show the [Fe/H] average values in 0.6\,Gyr $Age\_Param$ bins. Short vertical white lines account for the statistical error. Long vertical lines show the last star formation burst in \citet[][line colours and styles as in Fig.~\ref{fig:7}]{RuizLara2020}. Colour bars show the logarithm of the number of stars per [Fe/H]-age bin, normalised to unity.}
    \label{fig:9}
\end{figure*}

Regarding the evolution of [Fe/H], in Fig.~\ref{fig:9} we show that, as expected, the mean metallicity grows from almost primordial in the first stages of thin disk formation ($\sim$7\,Gyr) to almost solar in the present stages. For the N stars (left panel) the [Fe/H] variation is gently increasing, with small fluctuations. In the Y stars (right panel) we observe larger fluctuations. Both small variations in the N sample and large fluctuations in the Y sample coincide with the star formation events at $\sim$6\,Gyr and $\sim$1.5\,Gyr. This can be easily explained by the formation of a low-[Fe/H] population after Galactic-scale star formation events induced by external perturbations (see background colour contours in Fig.~\ref{fig:9}), which reduces the mean [Fe/H]. This low-[Fe/H] stellar population was born in galactic low-metallicity gas clouds after losing equilibrium due to an external perturbation.  Following this decrease, and just after the end of the star formation burst, metallicity increases again, both in N and Y stars (see the foreground colour plots in Fig.~\ref{fig:9}). This is also an expected behaviour after a strong star formation burst that is followed by intense episodes of stellar feedback. The larger variations observed in the Y stars might be due to the smaller size of the sample.

Variations in the metallicity of the star-forming gas also have a strong impact on the probability of the formation of planets and on their properties \citep{LivioPringle2003}. Low-metallicity Sun-like stars typically form with less massive accretion disks and host low-mass planets, while high-metallicity stars are born from massive accretion disks and are able to host giant planets \citep[][CH19]{Mulders2018,Santos2004,Mayor2011,Mortier2012,Buchhave2012,Johnson2012}. This also has an impact on the $^7$Li depletion rate during PMS \citep[][]{Eggenberger2012,Chavero2019,Cassisi2020} and on the strength of tidal interactions and the consequent planet engulfments \citep{AguileraGomez2016,Stephan2020,SoaresFurtado2020}. In the next paper from our collaboration we study the role that star-planet interaction plays in the generation of the old A(Li)-rich population observed in Fig.~\ref{fig:7} and its relation with galactic [Fe/H] evolution.

\begin{figure}
    \centering
    \includegraphics[width=0.49\textwidth]{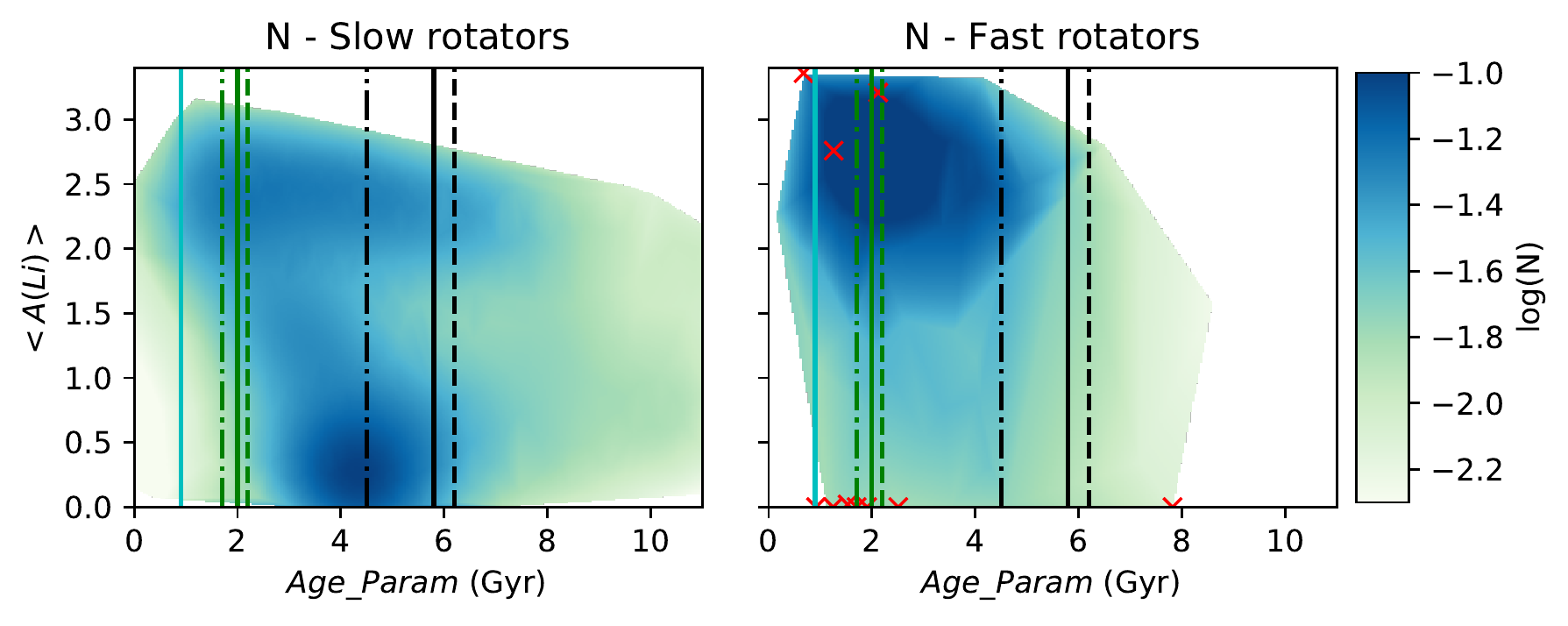}
    
    \includegraphics[width=0.49\textwidth]{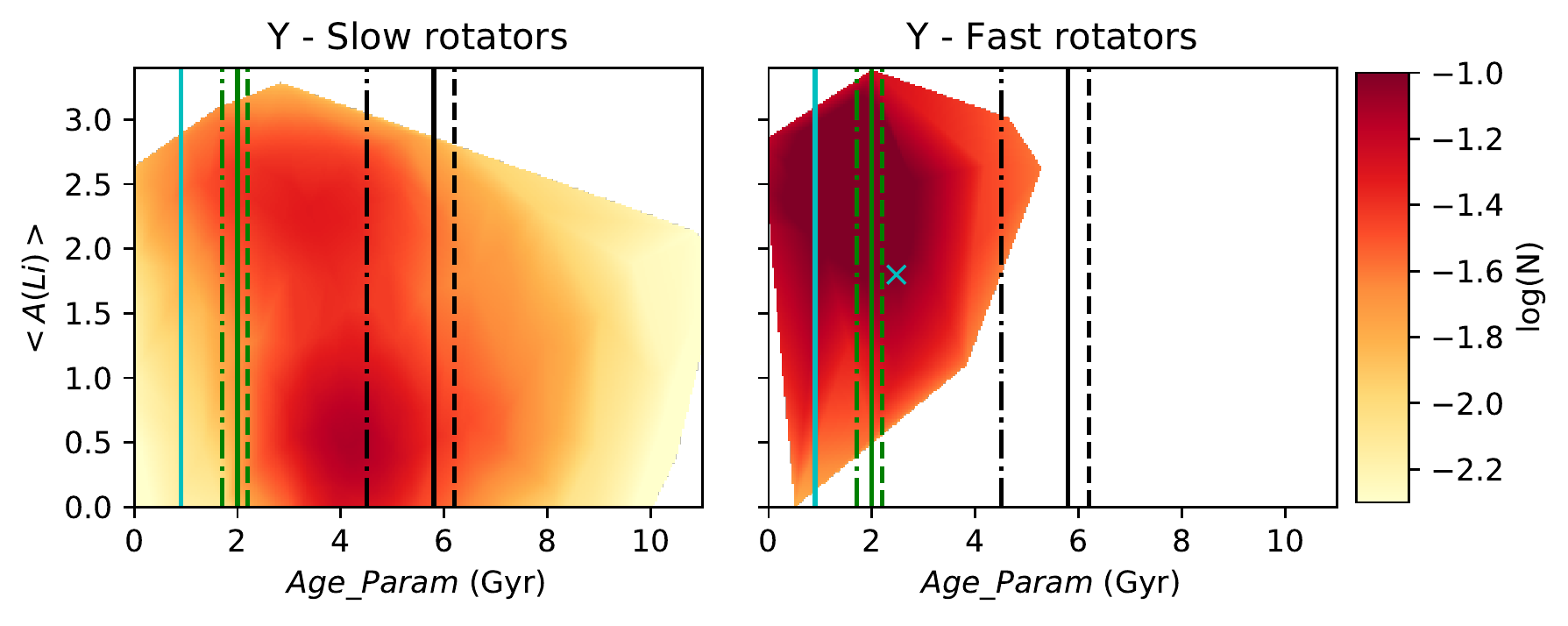}
    \caption{Like Fig.~\ref{fig:8}, but showing the contribution of slow rotators ($v\sin i <$ 8\,kms$^{-1}$, {\em left}) and fast rotators ($v\sin i >$ 8\,kms$^{-1}$, {\em right}) separately. Like in Fig.~\ref{fig:8}, colour bars show the logarithm of the number of stars per A(Li)-age bin, normalised to unity.}
    \label{fig:8b}
\end{figure}

Finally, although the effects of stellar rotation on the A(Li) depletion processes are still under debate, the stellar rotation evolution of FGK MS stars can add a new piece of information to help solve the general puzzle of galactic A(Li) evolution. It is relevant that the general behaviour of stellar rotation in FGK MS stars is to slow down with time \citep{vanSaders2013}. In Fig.~\ref{fig:8b} we show that the $v \sin i$ distribution of stars in our sample supports this hypothesis. We observe that the fast rotators (right panels) are mainly young A(Li)-rich stars, while slow rotators are slightly older and A(Li)-poorer. However, the presence of an old A(Li)-rich slow-rotator population in the N-star sub-sample is intriguing. From a first look this could be explained by intrinsic differences in the initial stellar rotation or by external effects, such as planet engulfment. However, it is well known that slow rotators are more efficient in depleting $^7$Li, so an initial slow rotation would lead to a fast A(Li) decrease \citep[CH19][]{ArencibiaSilva2020}; this is a fact that disproves the first hypothesis. Regarding the external effects, a planet in the process of spiralling onto the star boosts the turbulence in its convective zone. This process changes the properties of the convective layer, which has important consequences on the metal abundances in stellar atmospheres (including a A(Li) increase; \citealt{SoaresFurtado2020}) but only a minor impact on the angular momentum of the host star \citep{Stephan2020}. In our collaboration's next paper  we will present a detailed study of the [Fe/H], the rotation, and the planet engulfments connection with the galactic A(Li) evolution.

\section{Conclusions}\label{sec:conc}

In this paper we have analysed the A(Li) distribution of a new sample of FGK MS stars that belong to the MW thin disk. The main goal of this work is to explain the origin of the observed bi-modal A(Li) distribution in the A(Li)-$T_{\rm eff}$ plane and the origin of the isthmus region that is the `connector' of the two A(Li) populations, one rich, one poor. We aimed to demonstrate that this distribution is not a statistical artifice but a direct consequence of particular MW SFHs and GAHs, combined with the known $^7$Li production and depletion mechanisms.

After considering theory of galactic and stellar evolution, we developed a GCE model for $^7$Li that includes most of the known production and depletion  mechanisms. Comparing models with our data sample, we have confirmed that the SFH and the GAH play an important role in shaping the galactic A(Li) distribution. Furthermore, we have demonstrated that the A(Li) distribution can put new constrains on the SFH and GAH of galaxies. The general picture of the A(Li) evolution in thin-disk stars presented in this paper agrees well with recent investigations into the study of the MW SFH. As mentioned, we argue that our work supports the hypothesis that two to three star formation bursts, which were provoked by a satellite flyby ($\sim$0.5, 2.0, and 6\,Gyr ago), and a previous massive IGM gas inflow (6$-$7\,Gyr ago) shaped the chemical properties of stars in the galactic thin disk. A direct consequence of these Galactic-scale events is the creation of a bi-modal A(Li) distribution connected through a region we call the isthmus.

Finally, we have constructed the following general picture of the A(Li) evolution in the MW's thin-disk stars. This scenario is supported by results from our GCE model and from many other works that studied, among other things, the MW SFH, the $^7$Li depletion/production by several mechanisms, and stellar and planet formation and evolution. However, some of the hypothesis presented here need to be confirmed by data from future surveys and by improved GCE models:

\begin{itemize}
    \item Seven gigayears ago (and earlier): The MW contained only thick-disk stars, halo stars, and a small amount of low-[Fe/H] gas that was fuelling a low SFR \citep[][]{Haywood2016}. Metallicity of intergalactic gas was slowly increasing through the marginal SN feedback. The new stars had low but slightly increasing metallicity and a A(Li) similar to the one observed in the youngest thick-disk stars. Planet formation is not usual in metal-poor stars, so most stars did not develop planets \citep[e.g.][]{Johnson2012}. Stars with low metallicity and, thus, with no planets and/or massive protoplanetary disks could not slow down easily. As a consequence, they were mostly kept as fast rotators. The most efficient A(Li) depletion mechanism works only in stars with slow rotation \citep[CH19][]{ArencibiaSilva2020}. So, stars born from low-[Fe/H] gas, without planets, and with fast rotation depleted lithium only down to A(Li)$\sim$2.2 (the Spite plateau).
    \item About six to seven gigayears ago: The MW suffered a massive pristine gas inflow. This gas was settled to the thin disk and mixed with the slightly polluted gas already present therein \citep{Haywood2019}. This gas inflow was followed by a massive star formation event that was triggered by a satellite galaxy strongly interacting with the MW disk \citep{Mor2019}. The new stellar population inherited a low [Fe/H] and a A(Li) value close to primordial (2.7) from the ISM gas clouds.
    \item Three to six gigayears ago: Due to the strong star formation event that occurred about 6$-$7\,Gyr ago, the thin-disk gas was quickly polluted and new stars were born [Fe/H]-rich. These stars were able to develop massive protoplanetary disks and planets. The new [Fe/H]-rich stars (most of them with planets) became slow rotators and were able to efficiently deplete $^7$Li. Stars older than 3$-$4\,Gyr had enough time to deplete $^7$Li to a A(Li) value as low as 0 \citep[see][]{ArencibiaSilva2020}.
    \item From six gigayears ago to the present: New Galactic-scale events, both internal and external, have kept star formation ongoing. New stars have also formed from [Fe/H]-rich gas, becoming slow rotators and starting the $^7$Li depletion process. These stars have not yet had enough time to lower the A(Li) down to its lowest value. This is because the younger the stars are, the less the $^7$Li depletion they have suffered. This age-$^7$Li depletion dependence generated the feature we called the isthmus.
    \item From one to three gigayears ago: Two recent Galactic-scale events provoked a new star formation burst \citep{RuizLara2020}. New stars show a A(Li) that is similar to the one observed in newborn stars in the thin disk ($\sim3.2$).
\end{itemize}
In parallel to the aforementioned phenomena, another physical process has been shaping the A(Li) distribution. This mechanism has been producing a stellar population of A(Li)-rich slow rotators that span a wide range of ages (see Fig.~\ref{fig:8}). We propose that this process is either: the tidal interactions between the protoplanetary disk (or the young planets) and the host star, and the consequent planet engulfment; or a strong star-planet interaction \citep[see e.g.][]{SoaresFurtado2020}. Fully understanding the star-planet interactions is a challenging project that is not within the scope of this paper, but it is being studied by our research team as part of an ongoing project.

\begin{acknowledgements}
We thank the anonymous reviewer for his/her comments and suggestions that helped us improve this manuscript. SFR acknowledges support from a Spanish postdoctoral fellowship ‘Ayudas para la atracción del talento investigador. Modalidad 2: jóvenes investigadores, financiadas por la Comunidad de Madrid’ under grant number 2017-T2/TIC-5592. His work has been supported by the Madrid Government (Comunidad de Madrid-Spain) under the Multiannual Agreement with Complutense University in the line Program to Stimulate Research for Young Doctors in the context of the V PRICIT (Regional Programme of Research and Technological Innovation). SRF and CC acknowledge financial support from the Spanish Ministry of Economy and Competitiveness (MINECO) under grant number AYA2016-75808-R, AYA2017-90589-REDT, YA2016-79425-C3-2-P, and S2018/NMT-429, and from the CAM-UCM under grant number PR65/19-22462. FLA and R de la R. acknowledge support from the Faculty of the European Space Astronomy Centre (ESAC) - Funding references 569 and 570,  respectively. FLA would like to thank the technical support provided by A. Parras (CAB), Dr. J.A.Prieto (UCLM) and MSc J.Gómez-Malagón. 
This research has made use of the Spanish Virtual Observatory (http://svo.cab.inta-csic.es) supported from the Spanish MICINN/FEDER through grant AyA2017-84089. This research has made use of the NASA Exoplanet Archive, which is operated by the California Institute of Technology, under contract with the National Aeronautics and Space Administration under the Exoplanet Exploration Program.
\end{acknowledgements}

%
%

\begin{appendix} 
\section{The A(Li) bi-modality: K-S test}\label{app:KS}

To better characterise the two stellar populations (A(Li)-rich and A(Li)-poor stars), we used several simple K-S tests to analyse whether the two samples (N and Y stars) could be drawn from a uniform, Gaussian, or bi-Gaussian distribution. If the p value is below a certain significance level, typically 0.05, then the null hypothesis (i.e. the two samples can be drawn from a specific distribution) is rejected. We note that the p value is not the probability of the null hypothesis being true or false. In other words, a $p $ $value > 0.05$ does not mean that the two samples are strictly a bi-Gaussian distribution; it merely indicates that there is strong evidence that the two samples can be well described by it.\\

\begin{equation}
f(x)=\exp\left(-\frac{(x-\mu_1)^2}{(2\sigma_1)^2}\right)+a_{1,2}\exp\left(-\frac{(x-\mu_2)^2}{(2\sigma_2)^2}\right)
    \label{eq:2}
.\end{equation}

\begin{figure}
    \centering
    \includegraphics[width=0.45\textwidth]{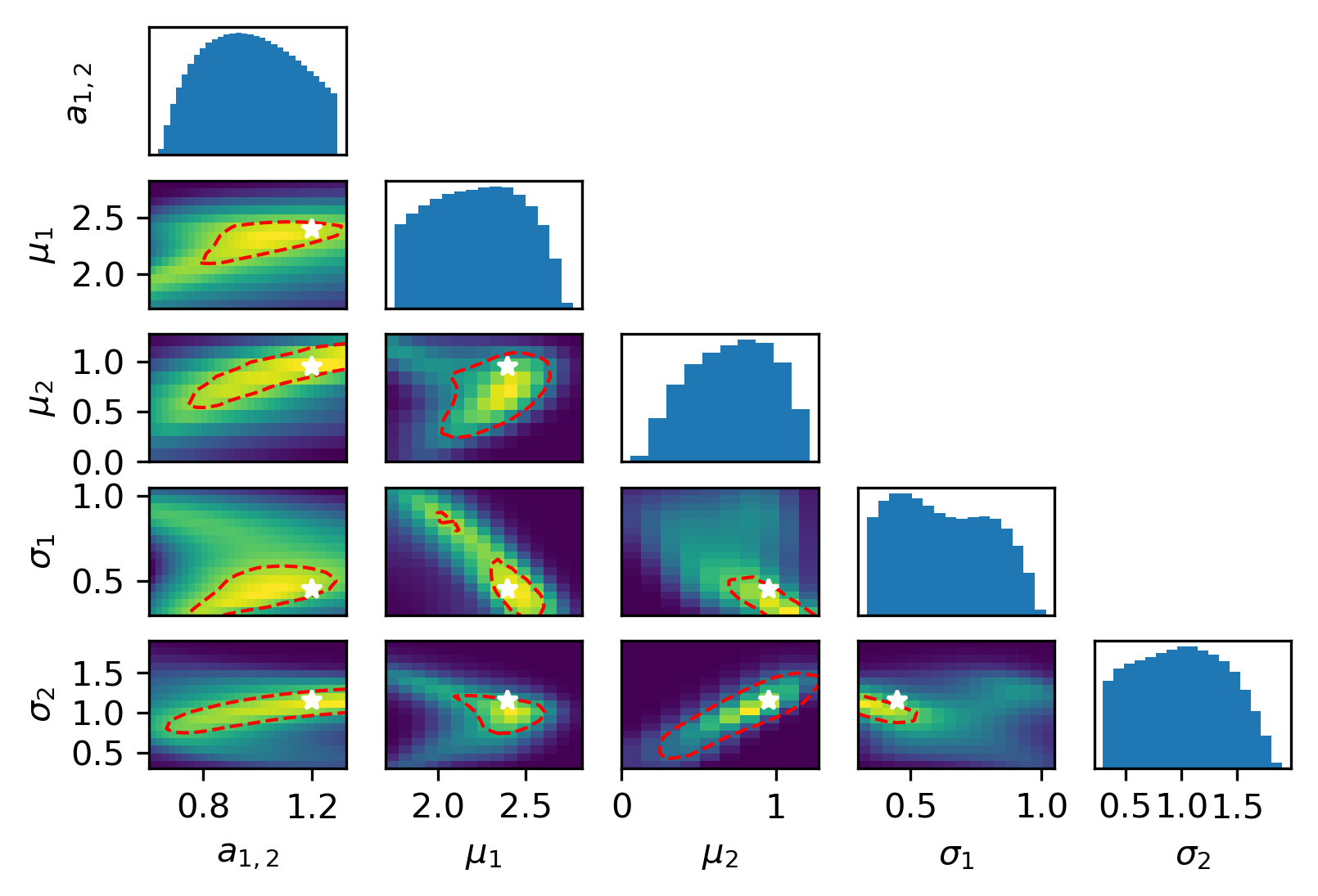}
    \caption{Corner plot showing the p values obtained from the K-S tests in the Y-star sample. We show the full set of p values obtained by scanning a wide range of bi-Gaussian distribution parameters (see Eq.~\ref{eq:2}). The red contour line encloses all p values above the 0.05 threshold. The white star shows the location of the highest p value.}
    \label{fig:B1}
\end{figure}

\begin{figure}
    \centering
    \includegraphics[width=0.45\textwidth]{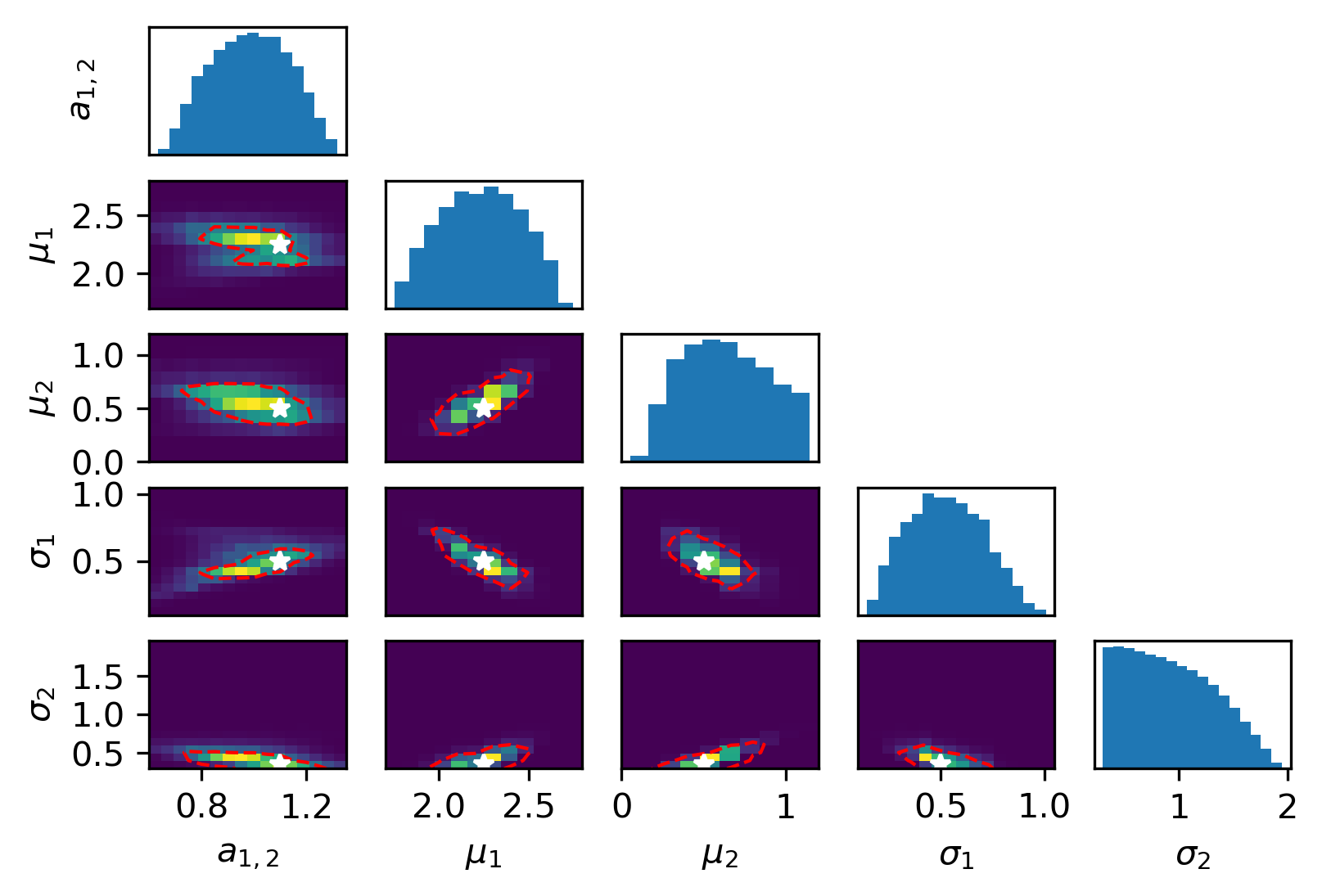}
    \caption{Analogous to Fig.~\ref{fig:B1}, but for N stars.}
    \label{fig:B2}
\end{figure}

In order to find the bi-Gaussian distribution that best describes the A(Li) data presented here, we used as the null hypothesis a bi-Gaussian distribution with five free parameters (see Eq.~\ref{eq:2}): the amplitude rate between the two Gaussian distributions, the two mean values, and the two standard deviations. It should be noted that we can recover a single Gaussian distribution by setting the amplitude rate to zero or by making the mean and standard deviation of both Gaussian distributions coincide. We can also recover a uniform distribution by setting a large enough value for the standard deviations. By scanning the full 5D parameter space, we were able to obtain the combination of free parameters that gives us the highest p value, for both the N and Y samples. For the sample of stars without known planets (N), we obtained: p value=0.11, $a_{1,2}=$1.1, $\mu_1=$2.25, $\mu_2=$0.5, $\sigma_1=$0.5, and $\sigma_2=$0.35. The results obtained for stars with detected planets (Y) are the following: p value=0.39, $a_{1,2}=$1.2, $\mu_1=$2.4, $\mu_2=$1.1, $\sigma_1=$0.3, and $\sigma_2=$1.1. The p values resulting from the full set of K-S tests are shown in Figs. \ref{fig:B1} and \ref{fig:B2}. The red contour line encloses the combination of parameters that results in a p value above the 0.05 limit. The white star indicates the location of the maximum p value in the corresponding parameter space.

\ 
Results obtained after applying the K-S test to both data samples (i.e. N and Y) allow us to reject the hypothesis that our data can be described by a single Gaussian (best p value =1.3$\times10^{-6}$) or a uniform distribution (best p value=2.8$\times10^{-6}$), for both the Y and N samples. \\

\section{$Age\_Lite$ and $Age\_Param$}\label{app:Ages}

In this work we analyse how several stellar parameters change with time. In particular, we present the histogram of the number of stars per age interval and the temporal evolution of global metallicity ([Fe/H]) and A(Li) as a function of stellar age. The age dependences we found allowed us to set a link between recently discovered Galactic-scale star formation events, several physical $^7$Li production and depletion mechanisms, and the A(Li) evolution.

It is clear that all the conclusions we reached from the analysis of these correlations are highly sensitive to the correctness of the age determination technique and, as such, its uncertainties. In order to prove the independence of our conclusions from the age scale, we present here figures that are equivalent to Figs.~\ref{fig:7}, \ref{fig:8}, and \ref{fig:9} except that they were created using $Age\_Lite$.

In Sect.~\ref{sec:observations} we fully describe the techniques we used to obtain the age on each age scale. $Age\_Param$ determination is based on isochrones fitting, while $Age\_Lite$ values were collected from recent literature.\\

\begin{figure*}
    \centering
    \includegraphics[width=0.45\textwidth]{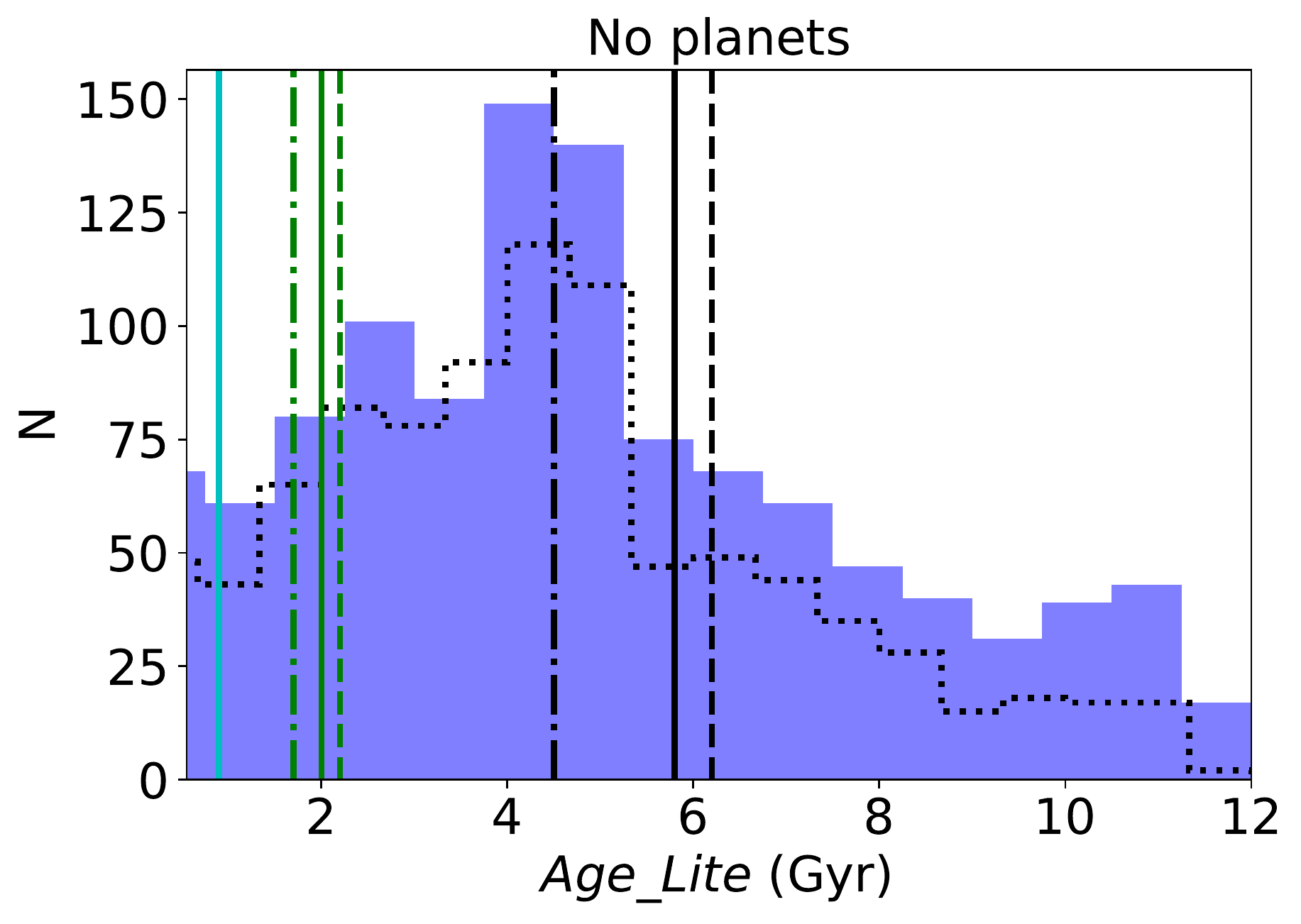}
    \includegraphics[width=0.45\textwidth]{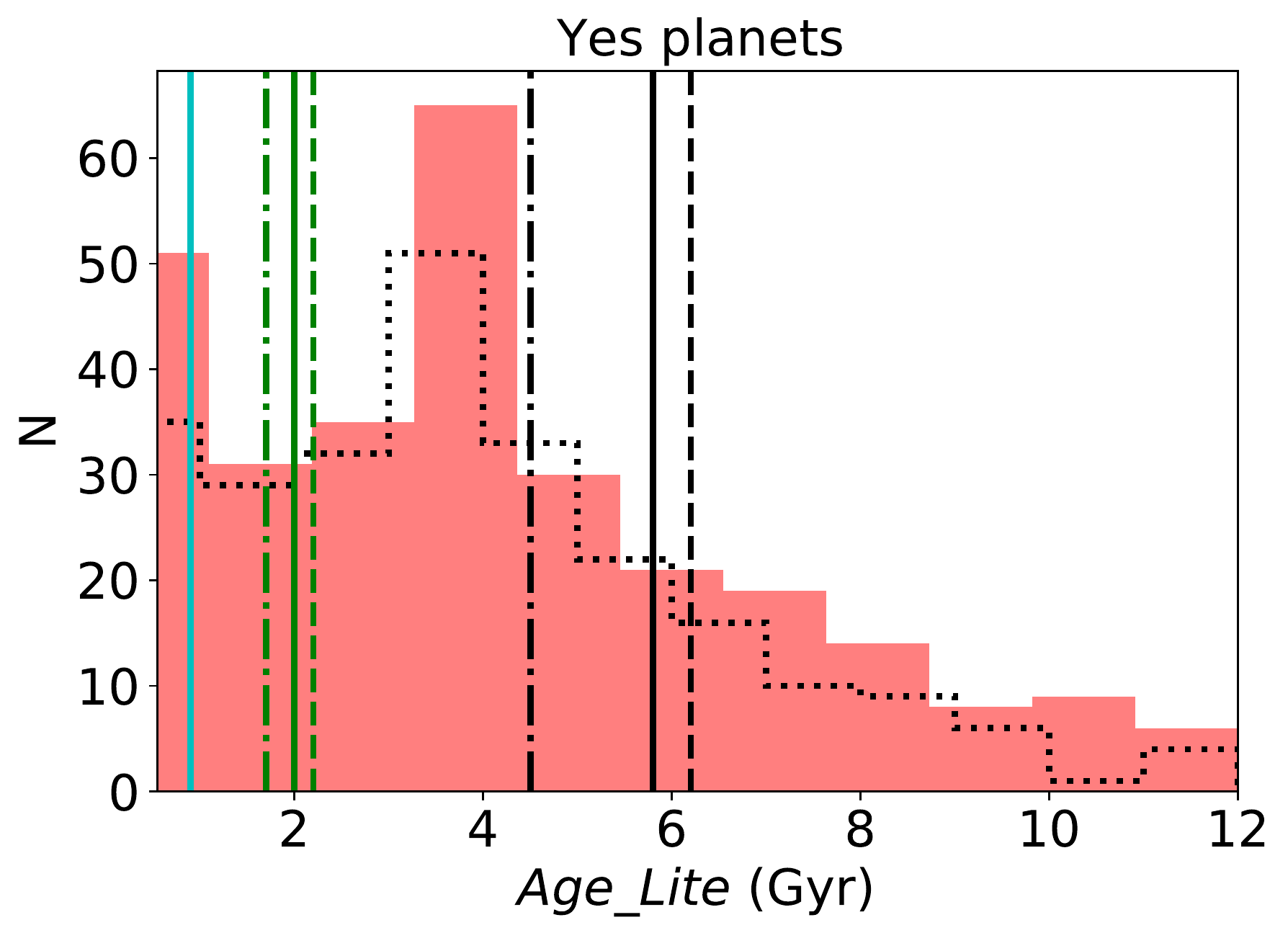}
    \caption{$Age\_Lite$ histogram for stars with ({\em right}) and without ({\em left}) planets. We show the thin disk contribution to the total with a black dotted histogram. The bin width has been obtained by applying the Freeman-Diaconis rule. Vertical lines show the start (dashed), peak (solid), and end (dot-dashed) of MW star formation bursts presented in \citet{RuizLara2020}: in black, the strong star formation burst detected in \citet{Mor2019}; in green, the last Sagittarius dwarf galaxy pericentre; and in cyan a very recent star formation burst also detected in the local stellar kinematics \citep{Antoja2018}.}
    \label{fig:A7}
\end{figure*}

\begin{figure*}
    \centering
    \includegraphics[width=0.45\textwidth]{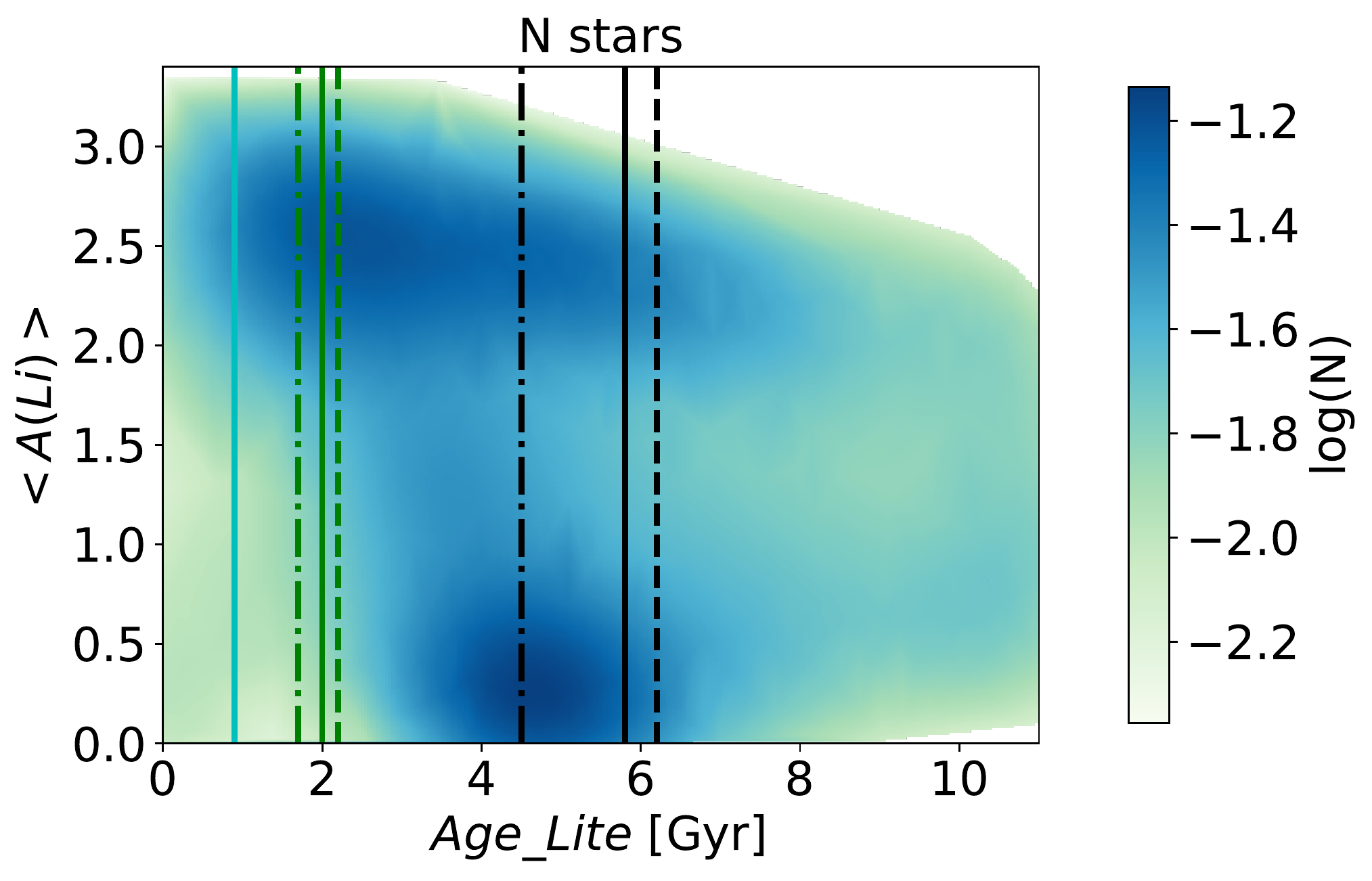}
    \includegraphics[width=0.45\textwidth]{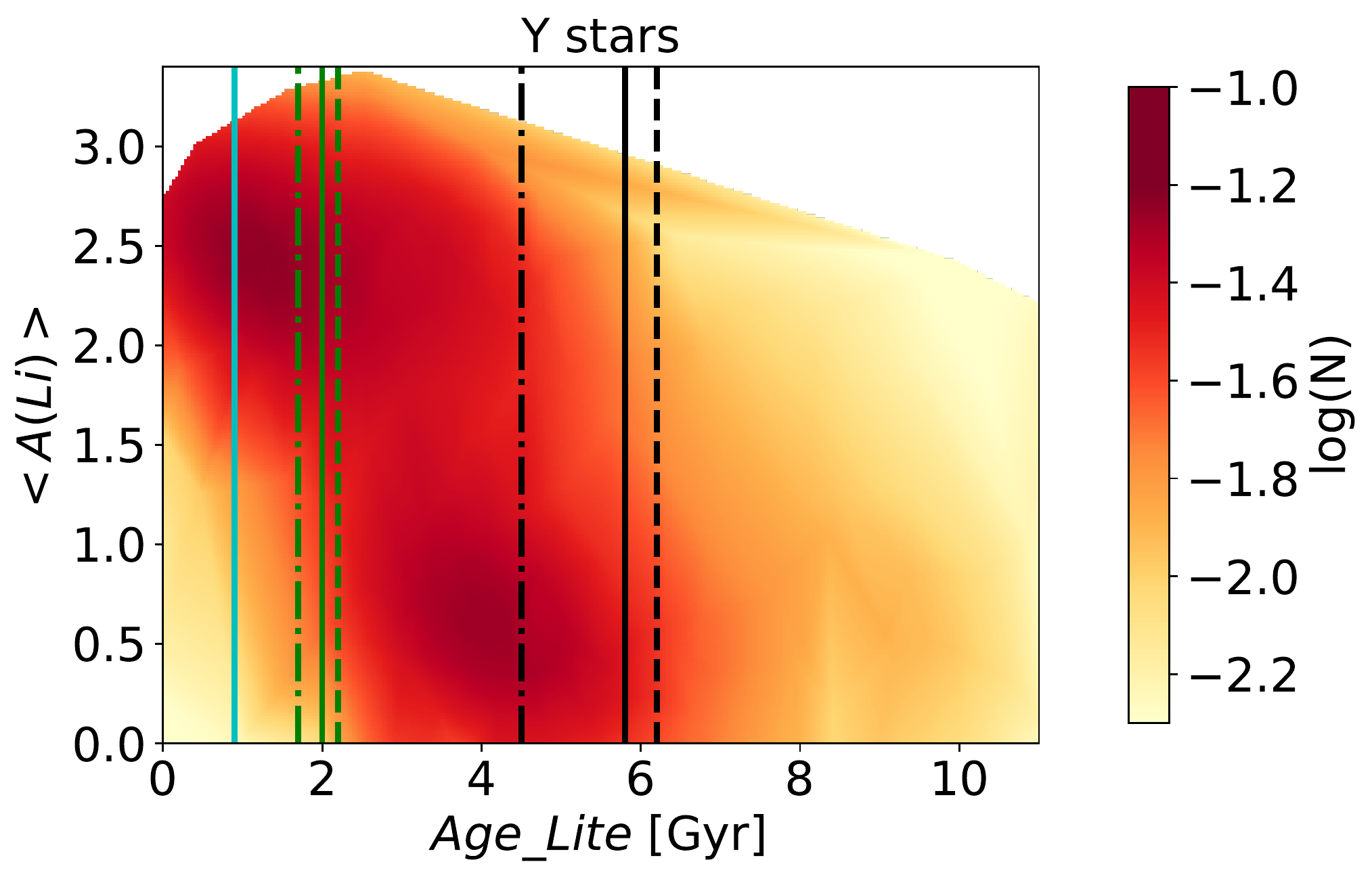}
    \caption{Mean A(Li) computed in 0.1 dex bins vs. mean $Age\_Lite$ in 0.6\,Gyr bins, for stars without known planets (N, {\em left}) and those that host confirmed planets (Y, {\em right}). Vertical lines show the last star formation burst in \citet[][line colours and styles as in Fig.~\ref{fig:7}]{RuizLara2020}. The colour bar shows the logarithm of the number of stars per A(Li)-age bin, normalised to unity. Equivalent plots with only thin- and/or thick-disk stars are available upon request to the authors.}
    \label{fig:A8}
\end{figure*}

\begin{figure*}
    \centering
    \includegraphics[width=0.45\textwidth]{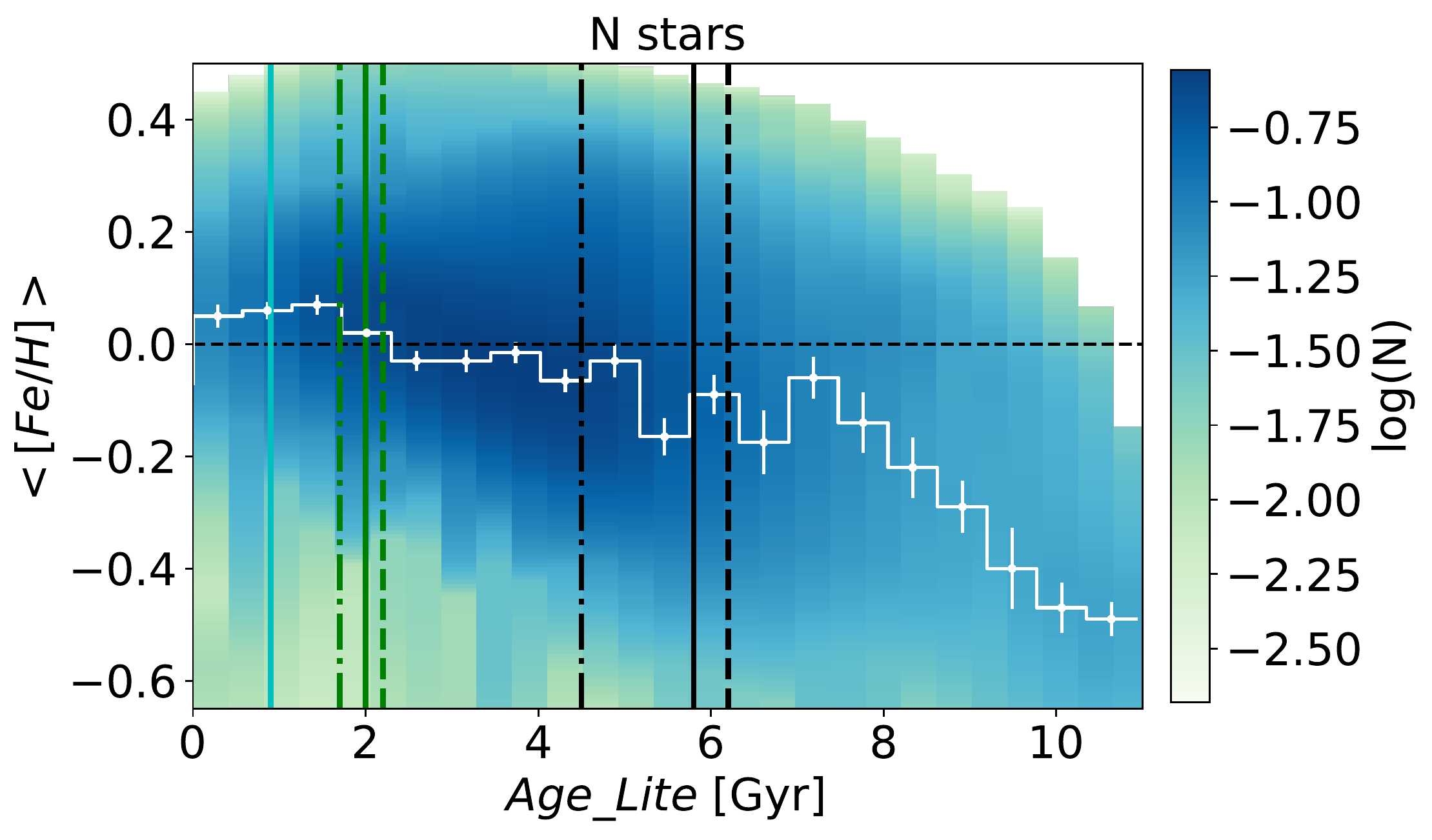}
    \includegraphics[width=0.45\textwidth]{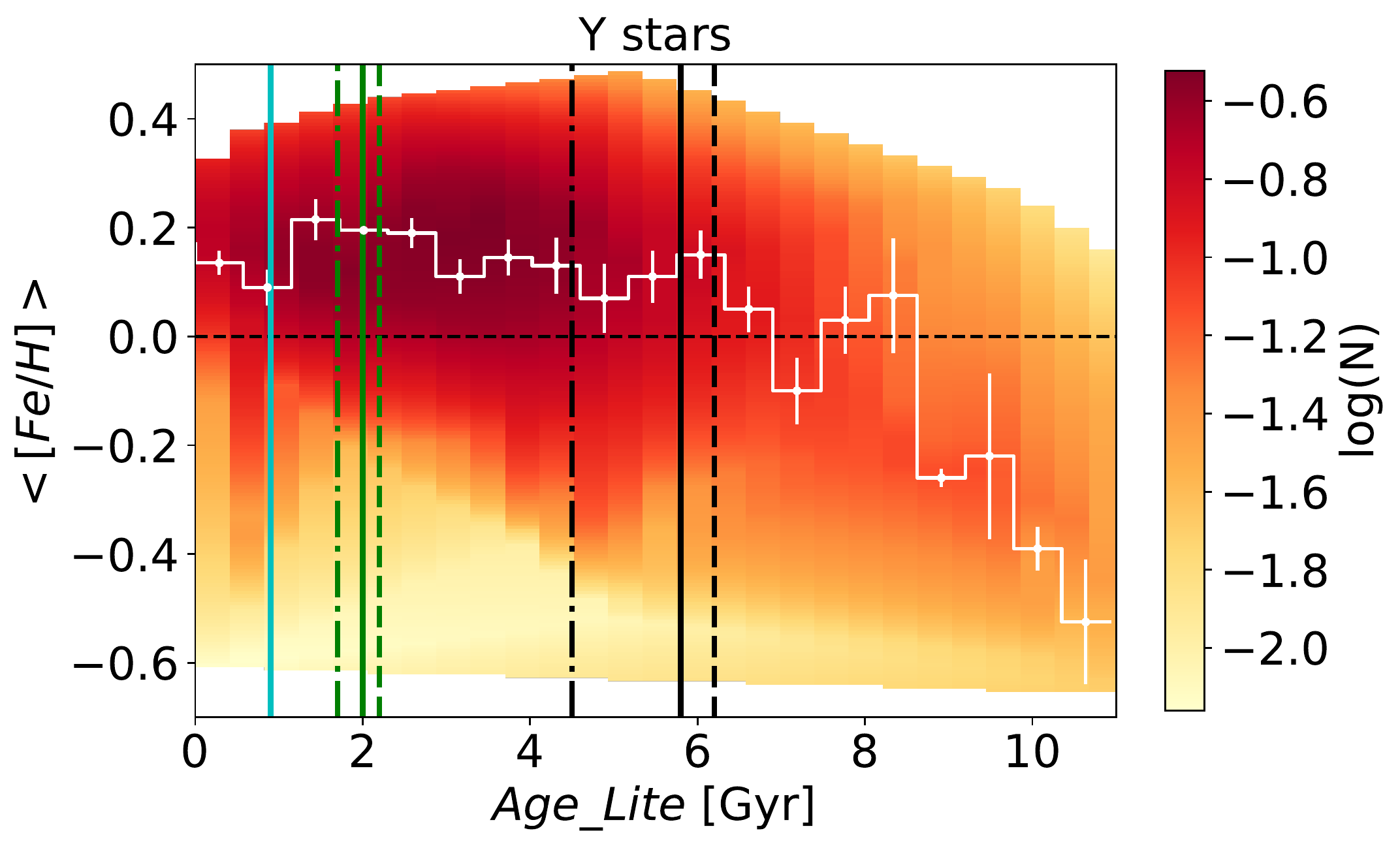}
    \caption{[Fe/H] vs. $Age\_Lite$ density plot, for stars without known planets (N, {\em left}) and those that host confirmed planets (Y, {\em right}). White solid lines show the [Fe/H] average values in 0.6\,Gyr $Age\_Param$ bins. Short vertical white lines account for the statistical error. Long vertical lines show the last star formation burst in \citet[][line colours and styles as in Fig.~\ref{fig:7}]{RuizLara2020}. Colour bars show the logarithm of the number of stars per [Fe/H]-age bin, normalised  to unity.}
    \label{fig:A9}
\end{figure*}

After confronting the $Age\_Lite$ figures (Figs.~\ref{fig:A7}, \ref{fig:A8}, and \ref{fig:A9}) with the analogous  $Age\_Param$ values (Figs.~\ref{fig:7}, \ref{fig:8}, and \ref{fig:9}), we conclude that both the global and local features mostly agree. First, the peaks in the star formation curves observed in Fig.~\ref{fig:7} coincide with those in Fig. \ref{fig:A7}, with differences close to the mean age uncertainties ($2$\,Gyr). These peaks in the star formation are correlated, in both figures, with the studied Galactic-scale events (see Sect.~\ref{sec:galactic}). Second, the A(Li) evolution and the groups detected in Fig.~\ref{fig:8} also appear in Fig.~\ref{fig:A8}, and almost at the same locations. Last, the [Fe/H] distribution and evolution in Fig.~\ref{fig:9}, and their relation with the Galactic-scale events, are also preserved when using the $Age\_Lite$ scale (Fig.~\ref{fig:A9} instead of $Age\_Param$).\\ \smallskip

\end{appendix}

\end{document}